\documentstyle[aps,epsf,rotate]{revtex}
\font\mb=msbm10

\begin{document}
\draft
\title{Simple deterministic dynamical systems with fractal diffusion
coefficients} 
\author{R. Klages$^{1,}$\cite{RKadd}, J.R. Dorfman$^2$}
\address{$^1$Institut f\"ur Theoretische Physik, Technische
Universit\"at Berlin, Hardenbergstra{\ss}e 36, D-10623 Berlin,
Germany; e-mail:R.Klages@physik.tu-berlin.de\\ 
$^2$Institute for Physical Science and Technology and Department
of Physics, University of Maryland, College Park,\\
MD 20742, USA; e-mail: jrd@ipst.umd.edu}
\date{\today}
\maketitle
\begin{abstract}
We analyze a simple model of deterministic diffusion. The model
consists of a one-dimensional periodic array of scatterers in which
point particles move from cell to cell as defined by a piecewise
linear map. The microscopic chaotic scattering process of the map can
be changed by a control parameter. This induces a parameter dependence
for the macroscopic diffusion coefficient. We calculate the diffusion
coefficent and the largest eigenmodes of the system by using Markov
partitions and by solving the eigenvalue problems of respective
topological transition matrices. For different boundary conditions we
find that the largest eigenmodes of the map match to the ones of the
simple phenomenological diffusion equation. Our main result is that
the difffusion coefficient exhibits a fractal structure by varying the
system parameter. To understand the origin of this fractal structure,
we give qualitative and quantitative arguments. These arguments relate
the sequence of oscillations in the strength of the
parameter-dependent diffusion coefficient to the microscopic coupling
of the single scatterers which changes by varying the control
parameter.
\end{abstract}

\section{Introduction}\label{txt:model}
Over the past years there has been a rapidly growing interest in
trying to understand the mechanism of nonequilibrium transport on the
basis of dynamical systems theory 
\cite{EvMo90,Hoov91,LiLi92,MaHo92,Wig92,Cvit95,Mare97,Gasp,DoVL}.
One line of work is related to computer simulations, where interacting
many-particle systems under nonequilibrium conditions, like shear
or an external field, are in the center of the investigations
\cite{PoHo87,PoHo88,PH89,ECM,DePH96}. Another line of research focuses
on low-dimensional models like the random
\cite{vBDo95,EDNJ95,LvBD97,DePo97} or periodic
\cite{MaZw83,MH87,GN,CvGS92,Vanc,Ch1,Gas93,BarEC,GaBa94,LNRM95,DeGP95,Gasp96,DeMo97} Lorentz gas. An even simpler model which shares
certain properties of the periodic Lorentz gas are two-dimensional
multibaker maps \cite{PG1,TG1,TG2,TeVB96,MoRo96,VTB97,GaKl,KlTe97}.
Lorentz gases and multibaker maps have become standard models in the
field of chaos and transport, since, on the one hand, they catch the
physical essence of certain real nonequilibrium processes, but, on the
other hand, they are still simple enough such that they can be
analyzed in detail theoretically. The final point along this line of
reduction of complexity is the problem of deterministic diffusion in
one-dimensional maps, as it has first been studied by Grossmann and
Fujisaka \cite{GF1,GF2,GrTh83}, by Geisel et al.\
\cite{GeNi82,GeTo84,GNZ85}, and by Schell, Fraser, and Kapral
\cite{SFK}. Later, methods of cycle expansion
\cite{Art1,ACL93,Art94,Tse94,Chen95} and other techniques have
been applied to this and related questions
\cite{Doer,Cla93,KlCP,ElKa,Kauf2}. The most interesting problem in this
context is to describe the dependence of deterministic diffusion on
varying a single parameter value. However, for computing
parameter-dependent diffusion coefficients so far either the
deterministic dynamics could not be treated in full detail, and thus
the results were approximative, or the task could be performed exactly
only for a few simple cases of parameter values.

The idea of this article is to apply methods of dynamical systems
theory, as discussed by Gaspard et al.\ \cite{Gas93,PG1,GB1}, to the
problem of parameter-dependent deterministic diffusion in
one-dimensional piecewise linear maps. In the remaining part of this
introductory section, we will explain the term deterministic
diffusion, and we will define the class of dynamical system we want to
analyze. In Section \ref{txt:fpm}, some important background of our
approach will be discussed briefly, and in Section
\ref{txt:sfp}, a method will be presented which enables the exact
computation of deterministic diffusion coefficients for a broad range
of parameter values. The result for the diffusion coefficient of the
simple map considered here turns out to be surprisingly complex so
that additional investigations, performed in Section \ref{txt:fdk},
are required to understand the origin of this unexpected non-trivial
diffusive behavior.

This article is based on the work of Ref.\ \cite{RKdiss}, Chapter 2, a
brief summary has been published in Ref.\ \cite{RKD}.

\subsection{What is deterministic diffusion?}
\begin{figure}
\epsfxsize=13cm
\centerline{\epsfbox{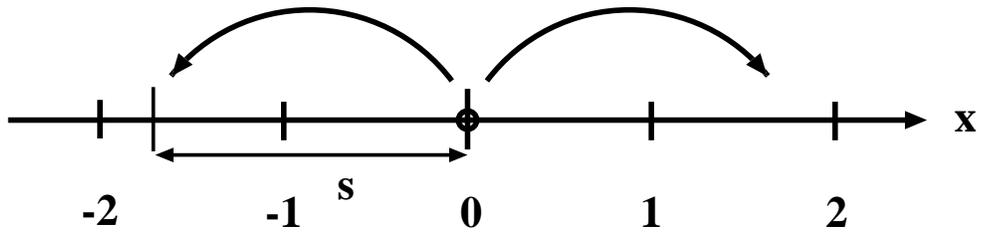}}
\caption{\label{fig:rwdif}Sketch of a simple one-dimensional random
walk model on the real line, in contrast to deterministic dynamics.}
\end{figure}

In a first approach, one may think about diffusion as a simple
non-correlated {\em random walk} \cite{Wax54,Wan66,vK,Reif87}, as
sketched in Fig.\ \ref{fig:rwdif}: One starts by choosing an initial
position $x_0$ of a point particle on the real line, e.g., $x_0=0$, as
shown in the figure. The dynamics of the particle is then determined
by fixing a probability density $\tilde{\rho}(s)$ such that
$p(s)=\tilde{\rho}(s)\, ds$ gives the transition probability $p(s)$
for the particle to travel from its old position $x_0$ over a distance
$s=x_1-x_0$ to its new position $x_1$. The same procedure can be
applied to any further change in the position of the particle. For
sake of simplicity, it is assumed that $\tilde{\rho}(s)$ is symmetric
with respect to $s=0$. The quantity of interest is here the orbit of
the moving particle, which is represented by a collection of points
$\{x_0,x_1,x_2,\ldots,x_n\}$ on the real line. $n$ refers to the
discrete time, which is given by the number of iterations, or
``jumps'', of the particle.

For performing a simple non-correlated random walk, it is assumed that
the probability density $\tilde{\rho}(s)$ is independent of the
position $x_n$ of the particle and of the discrete time $n$, i.e., it
is fixed in time and space. In other words, there is {\em no history}
in the dynamics of the particle, that is, the single iteration steps
are {\em statistically independent} from each other. This is called a
{\em Markov process} in statistical physics \cite{Wax54,vK,Beck}. Such
a random walk can be used as a simple model for diffusion: One starts
with an ensemble of particles at, e.g., $x_0=0$, and applies the
iteration procedure introduced above to each particle. If one
separates the real line into a number of subintervals of size $\Delta
x$, the number $n_p$ of particles in any such small interval after $n$
iterations can be counted. Divided by the total number $N_P$ of
particles, these results determine the probability density $\rho_n(x)$
for a sufficiently large number of particles. This quantity can be
interpreted as the probability to find one particle at a displacement
$x$ after $n$ iterations. Thus, $\rho_n(x)$ provides information about
the {\em macroscopic} distribution of an ensemble of particles, or of
the average position of one particle, respectively. The diffusion
coefficient $D$ can now be obtained from the second moment of the
probability density $\rho_n(x)$,
\begin{equation}
D=\lim_{n\to\infty}\frac{<x^2>}{2n} \quad , \label{eq:rwdk}
\end{equation}
where $<\ldots>:=\int dx \,\rho_n(x) \ldots$ represents the
probability density average. If the probability densities are
Gaussians, the diffusion coefficient of Eq.\ (\ref{eq:rwdk}) is
well-defined. This is the case for a {\em stochastic} diffusion
process as the one modeled here by a simple random walk.

In contrast to this traditional picture of diffusion as an
uncorrelated random walk, the theory of dynamical systems makes it
possible to treat diffusion as a {\em deterministic dynamical
process}: Here, the orbit of a point particle with initial condition
$x_0$ may be generated by a {\em chaotic dynamical system}
\begin{equation}
x_{n+1}=M(x_n) \quad . \label{eq:ddex}
\end{equation}
$M(x)$ is a one-dimensional map which determines how a particle gets
mapped from position $x_n$ to position $x_{n+1}$, as will be
introduced in detail below. The map $M(x)$ plays now the role of the
model-inherent probability density $\tilde{\rho}(s)$ of the previous
random walk: Defining $M(x)$ together with Eq.\ (\ref{eq:ddex}) gives
the full microscopic equations of motion of the system. Thus, the
decisive new fact which distinguishes this dynamical process from the
one of a simple uncorrelated random walk is that here $x_{n+1}$ is
{\em uniquely determined} by $x_n$ rather than having a distribution
of $x_{n+1}$ for a given $x_n$. This way, the {\em complete history}
of the particle is taken into account. If the resulting macroscopic
process of an ensemble of particles, governed by a deterministic
dynamical system like map $M(x)$, turns out to obey a law like Eq.\
(\ref{eq:rwdk}), i.e., if a diffusion coefficient exists for the
system, this process is denoted as {\em deterministic diffusion}
\cite{GF1,GF2,GrTh83,GeNi82,GeTo84,GNZ85,SFK,Art1,ACL93,Art94,Tse94,Chen95}.

\subsection{The deterministic model}\label{txt:formdef}
\begin{figure}
\epsfxsize=10.5cm
\centerline{\epsfbox{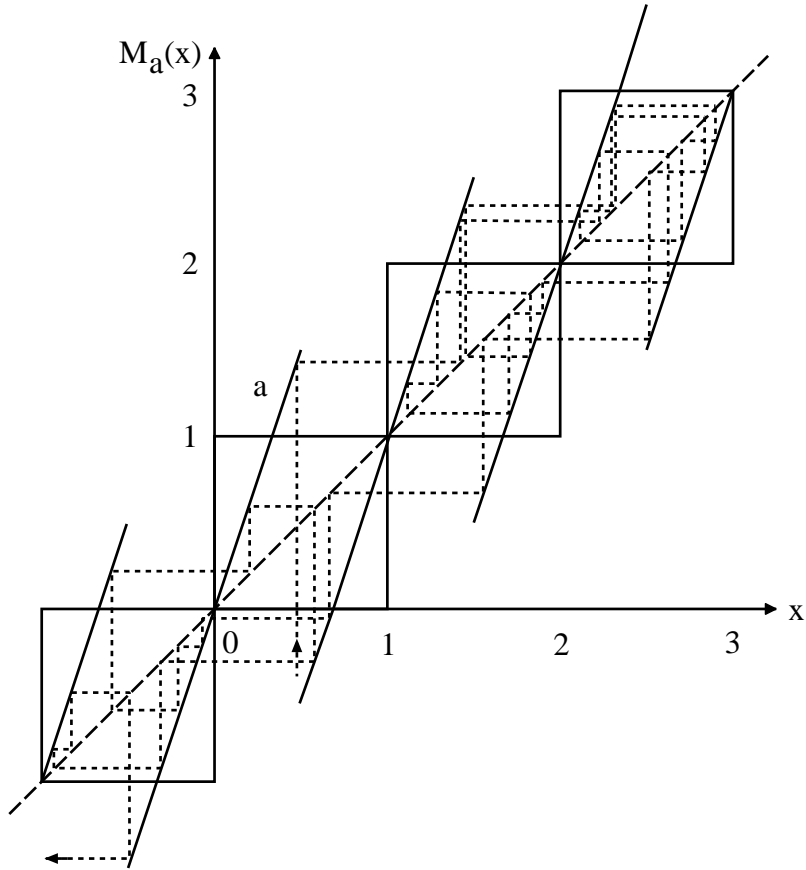}}
 
\vspace*{0.5cm}
\caption{\label{fig:model}Illustration of a simple model of
deterministic diffusion, see the dynamical system map $\cal L$,
Eqs.(\ref{eq:cob}) to (\ref{eq:mapa}), for the particular slope $a=3$.
The dashed line refers to the orbit of a moving particle. Its initial
condition is indicated by a black arrow close to the $x$-axis. The
particle moves under the action of the one-dimensional piecewise
linear map shown in the figure by jumping from box to box.}
\end{figure}

Fig.\ \ref{fig:model} shows the model which shall be studied in the
following. This model has apparently first been introduced by
Grossmann and Fujisaka \cite{GF1,GF2}. It depicts a ``chain of boxes''
which continues periodically in both directions to infinity, and the
orbit of a moving point particle. Let
\begin{equation}
M_a: \mbox{\mb R} \rightarrow \mbox{\mb R} \;\; , \;\; x_n \mapsto
M_a(x_n)=x_{n+1} \;\; , \;\; 
a>1\;\; ,\;\; x_n \:\in\: \mbox{\mb R}\;\; , \;\; n\: \in\: \mbox{\mb N}_0
\label{eq:cob}
\end{equation}
be a map modelling the chain of boxes introduced above, i.e., a {\em
periodic continuation} of {\em discrete one-dimensional piecewise
linear\cite{pili} expanding maps} with {\em uniform slope}. The index
$a$ denotes a control parameter, which is the absolute value of the
slope of the map, $x_n$ is the position of a point particle, and $n$
labels the discrete time. Since the map is expanding, i.e. $a>1$, its
Lyapunov exponent $\ln a$ is greater than zero. Thus, $M_a(x)$ is
dynamically instable and may in this sense be called chaotic
\cite{Ott}. In order for the map to be chaotic and piecewise linear,
it cannot be monotonic, so there must be points of discontinuity
and/or non-differentiability. The term ``chain'' in the
characterization of $M_a(x)$ can be made mathematically more precise
as a {\em lift of degree one},
\begin{equation} M_a(x+1)=M_a(x)+1 \quad , \label{eq:lft} \end{equation}
for which the acronym {\em old} has been introduced
\cite{M1,Als1,KaHa95}. This means that $M_a(x)$ is to a certain
extent translational invariant.  Being {\em old}, the full map
$M_a(x)$ is generated by the map of one box, e.g., on the unit
interval $0< x\le 1$, which will be referred to as the {\em box
map}. It shall be assumed that the graph of this box map is point
symmetric with respect to the center of the box at
$(x,y)=(0.5,0.5)$. This implies that the graph of the full map
$M_a(x)$ is anti-symmetric with respect to $x=0$,
\begin{equation}
M_a(x)=-M_a(-x)\quad , \quad \label{eq:sym}
\end{equation}
so that there is no ``drift'' in the chain of boxes. For convenience,
the class of maps defined by Eqs.\ (\ref{eq:cob}), (\ref{eq:lft}), and
(\ref{eq:sym}) shall be denoted as {\em class $\cal P$} (where $\cal
P$ stands for piecewise linear), and maps which fulfill the
requirements of class $\cal P$ shall be referred to as {\em class
$\cal P$-maps}. In Fig.\ \ref{fig:model}, which contains a section of
a simple class $\cal P$-map, the box map has been chosen to
\begin{equation}
M_a(x)  = \left\{ 
\begin{array}{r@{\quad,\quad}l}
a x & 0< x\le\frac{1}{2} \\ 
a x +1-a & \frac{1}{2} < x\le 1
\end{array}
\right\} \quad , \quad a\ge2 \label{eq:mapa} \quad ,
\end{equation}
cf.\ Refs.\ \cite{GF1,GB1,Ott}. This example can be best classified as
a {\em Lorenz map with escape} \cite{GH,PRS,GS,HS,Als2}.  The chaotic
dynamics of these maps is generated by a
``stretch-split-merge''-mechanism for a density of points on the real
line \cite{PRS}. As a class $\cal P$-map, Eq.(\ref{eq:mapa}), together
with Eqs.(\ref{eq:cob}), (\ref{eq:lft}), and (\ref{eq:sym}), will be
referred to as map $\cal L$. Other class $\cal P$-maps have been
considered in Refs.\ \cite{GF1,GF2,Art1,GB1,Gro3,Tho1}, another
example is discussed in Refs.\ \cite{RKdiss,dcrc,dcdd}.

It has been proposed \cite{GF2} to look at the dynamics in this chain
of boxes in analogy to the process of {\em Brownian motion}
\cite{Wax54,Wan66,vK}: If a particle stays in a box for a few
iterations, its {\em internal} box motion is supposed to get
randomized and may resemble the microscopic fluctuations of a Brownian
particle, whereas its {\em external} jumps between the boxes could be
interpreted as sudden ``kicks'' the particle suffers by some strong
collision. This suggests that ``jumps between boxes'' contribute most
to the actual value of the diffusion coefficient. Brownian motion is
usually described in statistical physics by introducing some
stochasticity into the equations which model a diffusion process. The
main advantage of the simple model discussed here is that diffusion
can be treated by taking the full dynamics of the system into account,
i.e., the {\em complete orbit} of the moving particle is
considered, without any additional approximations. This is another way
to understand the notion of deterministic diffusion in contrast to
diffusion as obtained from stochastic approaches. That is, in the
purely deterministic case the orbit of the particle is
immediately fixed by determining its initial condition.

One should note that the strength of diffusion, and therefore the
magnitude of the diffusion coefficient, are related to the probability
of the particle to escape out of a box, i.e., to perform a jump into
another box. This escape probability, however, as well as the average
distance a particle travels by performing such a jump, changes by
varying the system parameter. The problem which will be solved in the
following is to develop a general method for computing
parameter-dependent diffusion coefficients $D(a)$ for class $\cal
P$-maps. Here, map $\cal L$ will serve as a simple example. However,
the methods to be presented should work as well for any other
class $\cal P$-map, supposedly with analogous results (see, e.g.,
\cite{RKdiss,dcrc}.

\section{First passage method}\label{txt:fpm}
The methodology of first passage, as it has been developed in the
framework of statistical physics \cite{Wax54,vK}, deals with the
calculation of decay- or escape rates for ensembles of statistical
systems with certain boundary conditions. In recent work by Gaspard et
al., these methods have successfully been applied to the theory of
dynamical systems \cite{GN,Gas93,PG1,GB1,GD1,GD2}. In the following,
the principles of first passage for the class $\cal P$ of dynamical
systems defined above will be briefly outlined. The method will turn
out to provide a convenient starting point for computing
parameter-dependent diffusion coefficients.

One may distinguish three different steps in applying the method:

\underline{Step 1}: Solve the one-dimensional phenomenological {\em diffusion
equation}
\begin{equation}
\frac{\partial n}{\partial t}=D\:\frac{\partial^2n}{\partial x^2} \label{eq:f2}
\end{equation}
with suitable boundary conditions, where $n:=n(x,t)$ stands for the
macroscopic density of particles at point $x$ and time $t$.  This
equation serves here as a {\em definition} for the diffusion
coefficient $D$.

\underline{Step 2}: Solve the {\em Frobenius-Perron equation}
\begin{equation}
\rho_{n+1}(x)=\int dy \: \rho_n(y) \:\delta(x-M_a(y)) \quad , \label{eq:fppur}
\end{equation}
which represents the continuity equation for the probability density
$\rho_n(x)$ of the dynamical system $M_a(y)$ \cite{Ott,LaMa}.

\underline{Step 3}: For a chain of boxes of chainlength $L$, consider
the limit chainlength $L$ and time $n$ to infinity: {\em If} for given
slope $a$ the respective largest eigenmodes of $n$ and $\rho$ turn out
to be identical in an appropriate scaling limit, {\em then} $D(a)$ can
be computed by matching the eigenmodes of the probability density
$\rho$ to the particle density $n$: For {\em periodic boundary
conditions}, i.e., $n(0,t)=n(L,t)$ and $\rho_n(0)=\rho_n(L)$, one
obtains
\begin{equation}
D(a)=\lim_{L\to\infty}
\left(\frac{L}{2\pi}\right)^2 \gamma_{dec}(a) \quad , \label{eq:erf1}
\end{equation}
where $\gamma_{dec}(a)$ is the decay rate in the closed system to be
calculated directly from the Frobenius-Perron equation and therefore
determined by quantities of the deterministic dynamical system. For
{\em absorbing boundary conditions}, i.e., $ n(0,t)=n(L,t)=0$ and
$\rho_n(0)=\rho_n(L)=0$, the same procedure leads to
\begin{equation}
D(a)=\lim_{L\to\infty}
\left(\frac{L}{\pi}\right)^2 \gamma_{esc}(a) \quad , \label{eq:erf2}
\end{equation}
where $\gamma_{esc}(a)$ is the escape rate for the open system. This
quantity can be further determined by the escape rate formalism
\cite{DoVL,ER} to
\begin{equation}
\gamma_{esc}(a)=\lambda({\cal R} ;a)-h_{KS}({\cal R} ;a) \quad ,
\label{eq:erf3}
\end{equation}
where the Lyapunov exponent $\lambda({\cal R};a)$ and the
Kolmogorov-Sinai (KS) entropy $h_{KS}({\cal R};a)$ are defined on the
repeller $\cal R$ of the dynamical system. This equation is an
extension of Pesin's formula to open systems, which is obtained in
case of $\gamma_{esc}=0$. Eqs.(\ref{eq:erf1}) and (\ref{eq:erf2}) have
been applied to a variety of models, like the periodic Lorentz gas,
two-dimensional multibaker maps and certain one-dimensional chains of
maps, by Gaspard and coworkers \cite{PG1,GB1}. Eq.(\ref{eq:erf2}),
together with Eq.(\ref{eq:erf3}), has first been presented for the
two-dimensional periodic Lorentz gas \cite{GN} and has later been
generalized to other transport coefficients and dynamical systems
\cite{GD1,GD2}. However, although of fundamental physical importance,
it seems in general to be difficult to use this equation for practical
evaluations of $D(a)$, because usually the KS entropy is hard to
calculate \cite{Ott}. Instead, Eq.(\ref{eq:erf2}) with
Eq.(\ref{eq:erf3}) can be inverted to get the KS entropy via the decay
rate of the dynamical system of Eq.(\ref{eq:erf1}) to $ h_{KS}({\cal
R};a)=\lambda({\cal R};a)-\frac{1}{4}\gamma_{dec}(a)\quad
(L\rightarrow \infty) \quad , $ or by employing Eq.\ (\ref{eq:erf1})
via the diffusion coefficient in the limit of large $L$.

\section{Solution of the Frobenius-Perron equation}\label{txt:sfp}
Following the first passage method, the problem of computing
parameter-dependent diffusion coefficients essentially reduces to
solving the Frobenius-Perron equation for the dynamical system in a
certain limit. In this section, a general method will be presented by
which this goal can be achieved. Its principles will be illustrated by
performing analytical calculations for some special cases of
parameters of map $\cal L$. Our method is based on finding Markov
partitions and on defining respective transition matrices. This
approach is quite well-known, especially in the mathematical
literature \cite{Sin68a,Sin68b,Bow1,Rue78,Bow79,CFS82,Sin89,Rue} and
has been employed by many authors for the calculation of dynamical
systems quantities
\cite{Beck,GH,MoSe75,Boy1,FrBo81,BGB90,ElKa85,Gra88,Mac94,BST,MacK}. We
apply it here for the first time to compute the full
parameter-dependent deterministic diffusion coefficient.

\subsection{Transition matrix method}\label{txt:tmm}
\begin{figure}
\epsfxsize=7cm
\centerline{\epsfbox{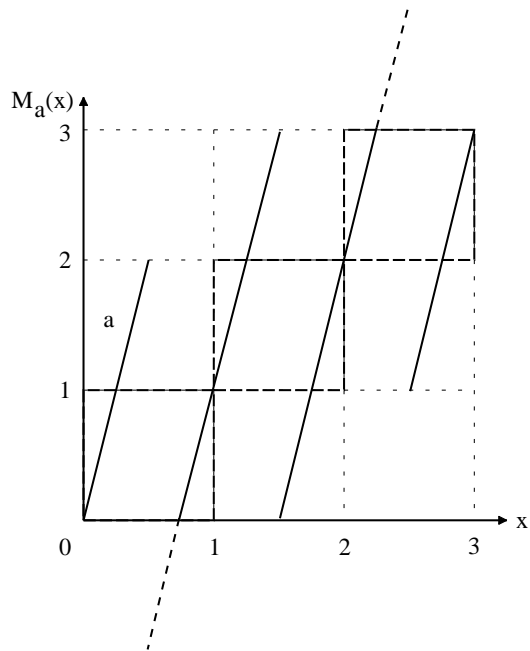}}
 
\vspace*{0.5cm}
\caption{\label{fig:mp4}Partition of map $\cal L$ at slope $a=4$
(dashed grid).}
\end{figure}

As a first example, the diffusion coefficient $D(a)$ shall be computed
for map $\cal L$ at slope $a=4$, as sketched in Fig.\ \ref{fig:mp4},
supplemented by periodic boundary conditions. The calculation will be
done according to the three-step procedure outlined above.

\underline{Step 1}: The one-dimensional diffusion equation Eq.(\ref{eq:f2}) can
be solved with periodic boundary conditions straightforward to
\begin{equation}
n(x,t)=a_0+\sum_{m=1}^{\infty}\exp\left(-(\frac{2\pi m}{L})^2Dt\right)
\left(a_m\cos(\frac{2\pi m}{L}x)+b_m\sin(\frac{2\pi m}{L}x)\right) \quad
, \label{eq:f2s}
\end{equation}
where $a_0\;,\;a_m$ and $b_m$ are the Fourier coefficients to be
determined by an initial particle density $n(x,0)$.

\underline{Step 2}: To solve the Frobenius-Perron equation, the key idea is
to write this equation as a matrix equation \cite{PG1,Beck}.  For this
purpose, one needs to find a suitable {\em partition} of the map,
i.e., a decomposition of the real line into a set of subintervals,
called {\em elements}, or {\em parts} of the partition. The single
parts of the partition have to be such that they do not overlap except
at boundary points, which are referred to as {\em points of the
partition}, and that they cover the real line completely \cite{Beck}.
In case of slope $a=4$, such a partition is naturally provided by the
box boundaries. The grid of dashed lines in Fig.\ \ref{fig:mp4}
represents a two-dimensional image of the one-dimensional partition
introduced above, which is generated by the application of the map.

Now an initial density of points shall be considered which covers,
e.g., the interval in the second box of Fig.\ \ref{fig:mp4}
uniformly. By applying the map, one observes that points of this
interval get mapped two-fold on the interval in the second box again,
but that there is also escape from this box which covers the third and
the first box intervals, respectively. Since map $\cal L$ is {\em
old}, this mechanism applies to any box of the chain of chainlength
$L$, modified only by the boundary conditions. Taking into account the
stretching of the density by the slope $a$ at each iteration, this
leads to a matrix equation of
\begin{equation}
\mbox{\boldmath $\rho$}_{n+1}=\frac{1}{a} \, T(a) \, \mbox{\boldmath
$\rho$}_n \quad , \label{eq:fpm}
\end{equation}
where for $a=4$ the $L$ x $L$-transition matrix $T(4)$ can be constructed
to
\begin{equation}
T(4)=
\left( \begin{array}{cccccccc} 2 & 1 & 0 & 0 & \cdots &0 & 0 & 1 \\
1 & 2 & 1 & 0 & 0 & \cdots & 0 & 0 \\
0 & 1 & 2 & 1 & 0 & 0 & \cdots & 0 \\
\vdots & & & \vdots & \vdots & & & \vdots \\
0 & \cdots & 0 & 0 & 1 & 2 & 1 & 0 \\
0 & 0 & \cdots & 0 & 0 & 1 & 2 & 1 \\
1 & 0 & 0 & \cdots & 0 & 0 & 2 & 1 \\ 
\end{array} \right) \quad . \label{eq:tm4}
\end{equation}
The matrix elements in the upper right and lower left edges are due to
periodic boundary conditions and reflect the motion of points from
the $L$th box of the chain to the first one and {\em vice versa}.

In Eq.(\ref{eq:fpm}), the transition matrix $T(a)$ is applied to a
column vector $\mbox{\boldmath $\rho$}_n$ of the probability density
$\rho_n(x)$ which, in case of $a=4$, can be written as
\begin{equation}
\mbox{\boldmath $\rho$}_n\equiv|\rho_n(x)>:=(\rho_n^1,\rho_n^2,\ldots,
\rho_n^k,\ldots,\rho_n^L)^*\quad , \label{eq:pdvec}
\end{equation}
where ``$*$'' denotes the transpose, and $\rho_n^k$ represents the
component of the probability density in the $k$th box,
$\rho_n(x)=\rho_n^k\;,\; k-1< x\le k\;,\;k=1,\ldots,L\;$, $\rho_n^k$
being constant on each part of the partition.

In case of $a=4$, the transition matrix is symmetric and can
be diagonalized by spectral decomposition. Solving the
eigenvalue problem
\begin{equation}
T(4)\,|\phi_m(x)>=\chi_m(4)\,|\phi_m(x)>\quad ,
\end{equation}
where $\chi_m(4)$ and $|\phi_m(x)>$ are the eigenvalues and
eigenvectors of $T(4)$, respectively, one obtains
\begin{eqnarray}
|\rho_n(x)>&=&\frac{1}{4}\sum_{m=0}^{L-1}\chi_m(4)\,
|\phi_m(x)><\phi_m(x)|\rho_n(x)> \nonumber \\
&=& \sum_{m=0}^{L-1}\exp\left(-n\ln\frac{4}{\chi_m(4)}\right) \,
|\phi_m(x)><\phi_m(x)|\rho_0(x)> \quad , \label{eq:pd4}
\end{eqnarray}
where $|\rho_0(x)>$ is an initial probability density vector and $\ln
4$ is the Lyapunov exponent of the map. Note that the choice of
initial probability densities is restricted by this method to
functions which can be written in the vector form of
Eq.(\ref{eq:pdvec}). For matrices of the type of $T(4)$, it is
well-known how to solve their eigenvalue problems
\cite{BeKa52,Kow54,Dav79}. This is performed in App.\
\ref{app:dkpbc} for a more general case, which includes the example
under consideration. For slope $a=4$, one gets
\begin{eqnarray}
\chi_m(4)&=&2+2\cos\theta_m\quad , \quad \theta_m:=\frac{2\pi}{L}m \quad ,
\quad m=0,\ldots,L-1\quad ; \nonumber \\
|\phi_m(x)>&=&(\phi_m^1,\phi_m^2,\ldots,\phi_m^k,\ldots,\phi_m^L)^*\quad
, \phi_m^k=\tilde{a}_m\phi_{m,1}^k+\tilde{b}_{m}\phi^k_{m,2}\quad , \nonumber \\
& & \phi_{m,1}^k:=\cos\theta_m(k-1)\quad ,\quad
\phi_{m,2}^k:=\sin\theta_m(k-1)\quad ,\quad \nonumber \\
& & k=1,\ldots,L\quad ,\quad k-1< x\le k \label{eq:ewp4} 
\end{eqnarray}
with $\tilde{a}_m$ and $\tilde{b}_m$ to be fixed by suitable normalization
conditions.

\underline{Step 3}: To compute the diffusion coefficient $D(4)$, it remains
to match the first few largest eigenmodes of the diffusion equation to
the ones of the Frobenius-Perron equation: In the limit as the time
$t$ and the system size $L$ approach infinity, the particle density
$n(x,t)$, Eq.(\ref{eq:f2s}), in the diffusion equation becomes
\begin{equation}
n(x,t)\simeq const. + \exp\left(-(\frac{2\pi}{L})^2Dt\right)\left(A\cos
(\frac{2\pi}{L}x)+
B\sin(\frac{2\pi}{L}x)\right)\quad , \label{eq:f2sa}
\end{equation}
where the constant represents the uniform equilibrium density of the
equation.

Analogously, for discrete time $n$ and chainlength $L$ to infinity,
one obtains for the probability density $\rho_n(x)$ of the
Frobenius-Perron equation, Eq.(\ref{eq:pd4}) with Eq.(\ref{eq:ewp4}), 
\begin{eqnarray}
\rho_n(x)&\simeq& const.+\exp\left(-\gamma_{dec}(4)n\right)\left(\tilde{A}\cos
(\frac{2\pi}{L}(k-1))+
\tilde{B}\sin(\frac{2\pi}{L}(k-1))\right)\quad , \nonumber \\
& & k=1,\ldots,L \quad , \quad k-1< x \le k \label{eq:pd4a}
\end{eqnarray}
with a decay rate of
\begin{equation}
\gamma_{dec}(4)=\ln\frac{4}{2+2\cos\frac{2\pi}{L}} \label{eq:dec4}
\end{equation}
of the dynamical system, determined by the second largest eigenvalue
of the matrix $T(4)$, see Eq.(\ref{eq:ewp4}). Note that the largest
eigenvalue is equal to the slope of the map so that for the first term
in Eq.(\ref{eq:pd4a}) the exponential vanishes, and one obtains a
uniform equilibrium density. Apart from generic discretization
effects in the time and position variables, which may be neglected in
the limit of time to infinity and after a suitable spatial coarse
graining, the eigenmodes of Eqs.(\ref{eq:f2sa}) and (\ref{eq:pd4a})
match precisely so that, according to Eq.(\ref{eq:erf1}), the
diffusion coefficient $D(4)$ can be computed to
\begin{equation}
D(4)=(\frac{L}{2\pi})^2\gamma_{dec}(4)=\frac{1}{4}+{\cal O}(L^{-4})
\quad . \label{eq:dkr4}
\end{equation}
This result is identical to what is obtained from a simple random walk
model \cite{SFK,RKdiss,dcdd}. The procedure can be generalized
straightforward to all even integers values of the slope, as is shown
in App.\ \ref{app:dkpbc}, and leads to a parameter-dependent
diffusion coefficient of
\begin{equation}
D(a)=\frac{1}{24}(a-1)(a-2) \quad , \quad a=2k\quad , \quad
k\,\in\, N \quad , \label{eq:dkre}
\end{equation}
in agreement with results of Ref.\ \cite{GF2}.  

A slightly more complicated example is the case of slope $a=3$, see,
e.g., Fig.\ \ref{fig:model}, which will be treated in the following.
Analogously to the previous example, for $a=3$ a simple partition can
be constructed, the parts of which are all of length $1/2$. According
to this partition, a transition matrix $T(3)$ can be determined, given
schematically by
\begin{equation}
T(3)=
\left( \begin{array}{ccccccccc} 1 & 1 & 0 & 0 & \cdots &0 & 0 & 1 & 0 \\
1 & 1 & 0 & 1 & 0 & 0 & \cdots & 0 & 0 \\
1 & 0 & 1 & 1 & 0 & 0 & \cdots & 0 & 0 \\
0 & 0 & 1 & 1 & 0 & 1 & 0 & 0 & \cdots \\
0 & 0 & 1 & 0 & 1 & 1 & 0 & 0 & \cdots \\
\vdots & & & & \vdots & & & & \vdots \\
0 & 1 & 0 & 0 & \cdots & 0 & 0 & 1 & 1 \\ 
\end{array} \right) \quad . \label{eq:tm3}
\end{equation}
Note that, in contrast to the case of $a=4$, here the matrix is formed
by submatrix {\em blocks} which move periodically to the right every
two rows. Since the partition of $a=3$ is a bit more complicated than
for $a=4$, the blocks refer to the partition of each box, whereas the
shift again is related to the lift property of the {\em old} map. The
matrix $T(3)$ is not symmetric. However, the eigenvalue problem of
this matrix can still be solved analogously to the case of $a=4$ (see
App.\ \ref{app:dkpbc}). The spectrum of the matrix turns out to be
highly degenerate, and therefore $T(3)$ cannot be simply diagonalized
anymore \cite{ZuFa84}. This is due to the fact that the matrix $T(3)$
is {\em non-normal}, i.e., $T(3)T^*(3)\ne T^*(3)T(3)$, which means
that it does not provide a system of orthogonal eigenvectors. Of
course still a transformation onto Jordan normal form to
``block-diagonalize'' this matrix could be applied. However, this
seems to be more useful for proving mathematical theorems than for
analytical calculations of diffusion coefficients \cite{Kla95}. Since
the probability density of the Frobenius-Perron equation is determined
by iteration of transition matrices, cf.\ Eq.\ (\ref{eq:fpm}), for
slope $a=3$ it is not in advance clear how the single eigenmodes
``mix'' in the limit time and chainlength to infinity and whether they
can be matched to the solutions of the diffusion equation as before so
that the diffusion coefficient $D(3)$ is again simply determined by
the second largest eigenvalue of the matrix.

\begin{figure}
\epsfxsize=10cm
\centerline{\rotate[r]{\epsfbox{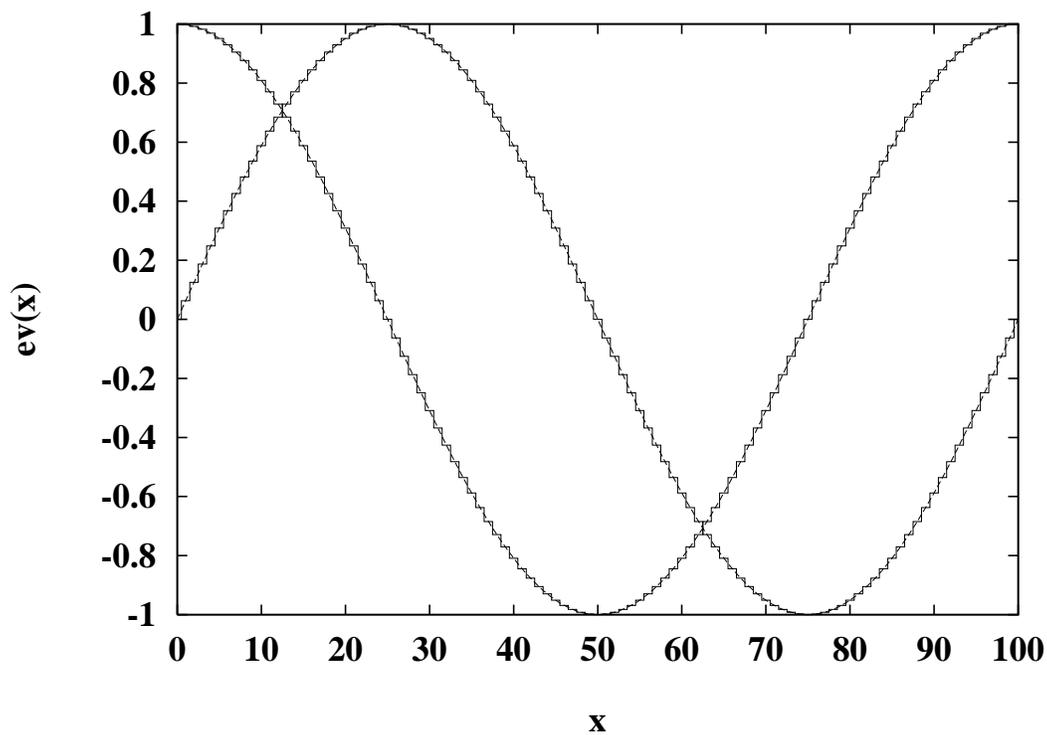}}}
 
\vspace*{0.5cm}
\caption{\label{fig:em3}The two second largest eigenmodes of map $\cal L$,
chainlength $L=100$, for slope $a=3$ with periodic boundary
conditions. The eigenmodes exhibit a step-like fine structure and
differ by a phase shift. In comparison, the respective two eigenmodes
obtained from solving the diffusion equation Eq.(\ref{eq:f2}) have
been included as dashed lines. They are almost indistinguishable from
the map eigenmodes.}
\end{figure}

This problem will first be approached pragmatically. Analogously to
the analytical solutions of Eqs.\ (\ref{eq:f2sa}) and (\ref{eq:pd4a})
for slope $a=4$, Fig.\ \ref{fig:em3} shows a plot of the two second
largest eigenmodes of $T(3)$ in comparison to the solution of the
diffusion equation. Again, one observes total agreement, except
differences in the fine structure. The same is true for the other
first few largest eigenmodes of $T(3)$. Thus, although straightforward
diagonalization and, therefore, a simple solution of the
Frobenius-Perron equation like Eq.\ (\ref{eq:pd4}) are not possible
anymore, the largest eigenmodes of $T(3)$ behave correctly in the
sense of the phenomenological diffusion equation so that it is
suggestive to compute the diffusion coefficient $D(3)$ via the second
largest eigenvalue of $T(3)$ again. With
\begin{equation}
\gamma(3)=\ln \frac{3}{1+2\cos(2\pi/L)} \quad ,
\end{equation}
see App.\ \ref{app:dkpbc}, and Eq.(\ref{eq:erf1}), one gets
\begin{equation}
D(3)=\frac{1}{3}+{\cal O}(L^{-4}) \quad . \label{eq:dkr3}
\end{equation}
As for $a=4$, this result is obtained as well from a simple random
walk model. However, to produce this value, the respective random walk
has to be defined in a slightly different way than for $a=4$
\cite{RKdiss,dcdd}.  Analogously to the case of even integer slopes, the
exact calculations can be generalized to all odd integer values of the
slope and lead to (see App.\ \ref{app:dkpbc})
\begin{equation}
D(a)=\frac{1}{24}(a^2-1) \quad , \quad a=2k-1 \quad , \quad k\,\in\,
N \quad , \label{eq:dkro}
\end{equation}
which again is identical to the result of Ref.\ \cite{GF2}.

Before this approach will be extended to other parameter values of the
slope in the following section, some details will be discussed which
are closely related to the calculations above: A special feature of
the diffusion coefficient results for integer slopes shall be pointed
out; further limits of this method with respect to generalized
diffusion coefficients shall be critically discussed; and the
application of the method to absorbing boundary conditions shall be
briefly outlined.

\subsubsection{Diffusion coefficients for integer slopes} 
Eqs.\ (\ref{eq:dkr4}) and (\ref{eq:dkr3}) show already that
$D(4)<D(3)$, which is at first sight counterintuitive. By evaluating
the general formulas of $D(a)$ given by Eqs.\ (\ref{eq:dkre}) and
(\ref{eq:dkro}) at other even and odd integer slopes, one realizes
that this inequality reflects a general oscillatory behavior of $D(a)$
at integer slopes. This result has already been obtained by Fujisaka
and Grossmann \cite{GF2}, but it has not been discussed in further
detail in their work.  A similar oscillatory behavior has been
observed for deterministic diffusion in certain classes of
two-dimensional maps
\cite{ReWi80,RRW81,DMP89,Dan89,Lebo}. This behavior cannot be understood
completely by one consistent simple random walk model
\cite{RKdiss,dcdd}.

\subsubsection{Matching lower eigenmodes} 
There appear serious problems in trying to extend the matching
eigenmodes procedure to arbitrarily low eigenmodes, even in case of
$a=4$, where the matrix is diagonalizable. With Eq.(\ref{eq:ewp4}),
one can check that
\begin{equation}
\phi_{m,1}^k=\phi_{L-m,1}^k \;\; , \;\;
\phi_{m,2}^k=-\phi_{L-m,2}^k\;\; ;
\;\; k=1,\ldots L \;\; , \;\; m=1,\ldots ,L-1 \; ,
\end{equation}
i.e., in contrast to the $m$ eigenmodes of the diffusion equation
Eq.(\ref{eq:f2}), the frequency of the eigenmodes of $T(4)$ does not
increase monotonically with $m$, but gets ``flipped over'' at the
($L/2$)th (for $L$ even) eigenmode such that the first and the last
half of the number of eigenmodes are identical, except a minus
sign. This is due to the discretization of the position variable $x$
in the diffusion equation to $k$ in the Frobenius-Perron {\em matrix}
equation Eq.(\ref{eq:fpm}), which was one of the basic ingredients for
the possibility to construct transition matrices.

Moreover, one should note that, according to Eq.(\ref{eq:ewp4}), the
smallest eigenvalue of $T(4)$ is equal to zero. For $T(3)$, a large
number of eigenvalues are even less than zero (see Eq.\
(\ref{eq:oddev}) in App.\ \ref{app:dkpbc} for $a=3$). Thus, except
for the first few largest eigenmodes, which still match reasonably
well to the eigenmodes of the diffusion equation in the limit time $n$
and chainlength $L$ to infinity, one cannot expect the method to work
simply that the components of a ``time-dependent'' diffusion
coefficient $D_n(a)$ are determined by smaller eigenvalues of the
transition matrices in straight analogy to Eq.(\ref{eq:erf1}).  This
could be taken as a hint that, to obtain more details of the dynamics,
refined methods are needed. For example, in Ref.\
\cite{PG3} the first orders of a position-dependent diffusion
coefficient have been determined for a class $\cal P$-map according to
a procedure which avoids the discretization of the real line.

\subsubsection{Absorbing and periodic boundary conditions}
The same procedure as outlined for periodic boundary conditions can
also be employed for absorbing boundaries. It shall be sketched
briefly, according to the three steps distinguished before:

\underline{Step 1}: The one-dimensional diffusion equation with absorbing
boundary conditions can be solved to
\begin{equation}
n(x,t)=\sum_{m=1}^{\infty}a_m\exp\left(-(\frac{\pi m}{L})^2Dt\right)
\sin(\frac{\pi m}{L}x)
\end{equation}
with $a_m$ denoting again the Fourier coefficients.

\begin{figure}
\epsfxsize=13cm
\centerline{\epsfbox{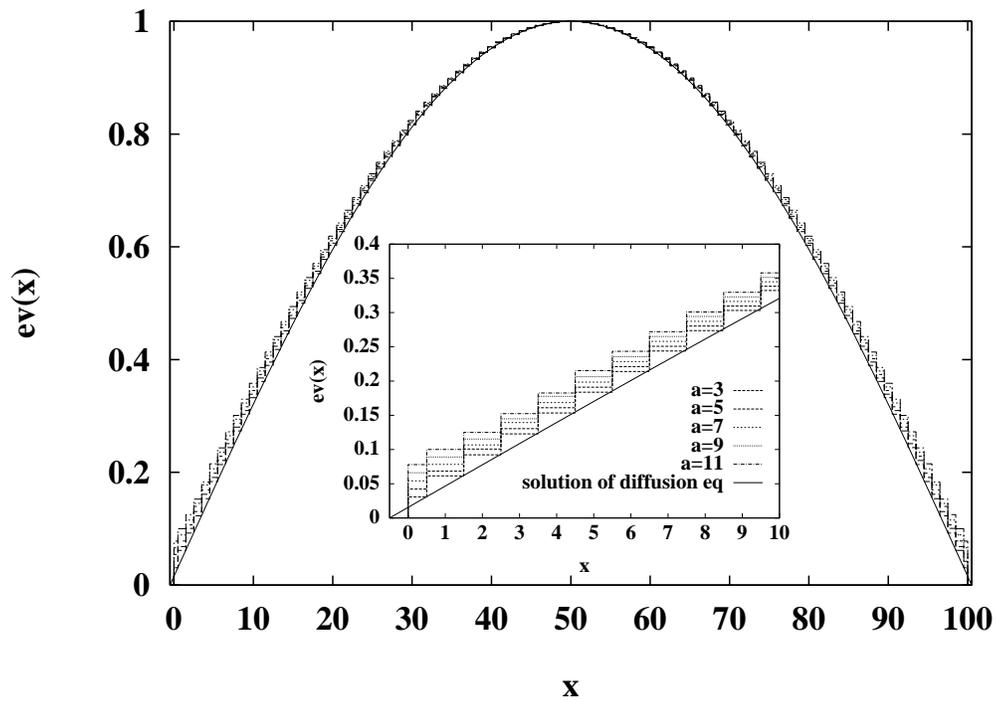}}
 
\vspace*{0.5cm}
\caption{\label{fig:emabc}Largest eigenmodes of map $\cal L$
for odd integer values of the slope $a$ with absorbing boundary
conditions and comparison to the largest eigenmode of the diffusion
equation Eq.(\ref{eq:f2}). The inset is a magnification of the region
around $x=0$.}
\end{figure}

\underline{Step 2}: The transition matrices for $a=4$ and $a=3$ at these
boundary conditions are identical to the ones of
Eqs.(\ref{eq:tm4}),(\ref{eq:tm3}), except that the matrices now
contain zeroes as matrix elements in the upper right and lower left
corners. However, due to this slight change in their basic structure
there is no general method to solve the eigenvalue problems for this
type of matrices anymore, in contrast to the case of periodic boundary
conditions. At least for $a=3$ and $a=4$, it is still possible to
obtain analytical solutions by straightforward calculations analogous
to the ones performed in Ref.\ \cite{PG1} (see App.\
\ref{app:dkabc}), but for any higher integer value of the slope even
these basic methods fail. This appears to be caused by strong boundary
layers. Fig.\ \ref{fig:emabc} shows numerical solutions for the
largest eigenmodes of the first odd integer slope transition matrices
in comparison to the solutions of the diffusion equation
Eq.(\ref{eq:f2}) (details of the numerics applied here are given in
the following section). It can be seen that near the boundaries, there
are pronounced deviations between the Frobenius-Perron and the
diffusion equation solutions. These deviations are getting smaller in
the interior region of the chain, but are gradually getting stronger
with increasing the value of the slope, as is shown in the
magnification. The same behavior can be found for even integer slopes,
although the quantitative deviation of these eigenmodes from the ones
of the diffusion equation solutions is slightly less than for odd
values of the slope. Thus, obviously absorbing boundary conditions
disturb the deterministic dynamics significantly, whereas similar
effects do not occur for periodic boundary conditions, which therefore
could be characterized as a kind of ``natural boundary conditions''
for this periodic dynamical system.

\underline{Step 3}: The different boundary conditions do not only show up in
the eigenmodes of the transition matrices, but also in the calculation
of the diffusion coefficients. In analogy to periodic boundary
conditions, the escape rate of the dynamical system at $a=3$ and $a=4$
is determined to
\begin{equation}
\gamma_{esc}(a)=\ln\frac{a}{\chi_{max}(a)} \label{eq:escabs}
\end{equation}
with $\chi_{max}(a)$ being the largest eigenvalue of the transition
matrix (see App.\ \ref{app:dkabc}),
\begin{equation}
\chi_{max}(3)=1+2\cos\frac{\pi}{L+2}\quad \mbox{and} \quad
\chi_{max}(4)=2+2\cos\frac{\pi}{L+1}\quad .
\end{equation}
Feeding this into Eq.(\ref{eq:erf2}) via matching eigenmodes, one
obtains
\begin{eqnarray}
D(3)&\simeq& \frac{1}{3}\frac{L^2}{(L+2)^2}+{\cal O}(L^{-4}) \;
\rightarrow \;
\frac{1}{3} \quad (L\rightarrow \infty)\quad , \nonumber \\
D(4)&\simeq& \frac{1}{4}\frac{L^2}{(L+1)^2}+{\cal O}(L^{-4}) \;
\rightarrow \;
\frac{1}{4} \quad (L\rightarrow \infty)\quad , \label{eq:dkl}
\end{eqnarray}
which gives a convergence of the diffusion coefficient with the
chainlength $L$ significantly below that obtained from periodic
boundary conditions. For example, for a chainlength of $L=100$ the
convergence is about two orders of magnitude worse. It can be
concluded that the transition matrix method works in principle for
absorbing boundary conditions as well, but that here its range of
application to compute diffusion coefficients is qualitatively and
quantitatively more restricted because of long-range boundary layers.

\subsection{Markov partitions}\label{txt:mp}
\begin{figure}
\epsfysize=20cm
\centerline{\epsfbox{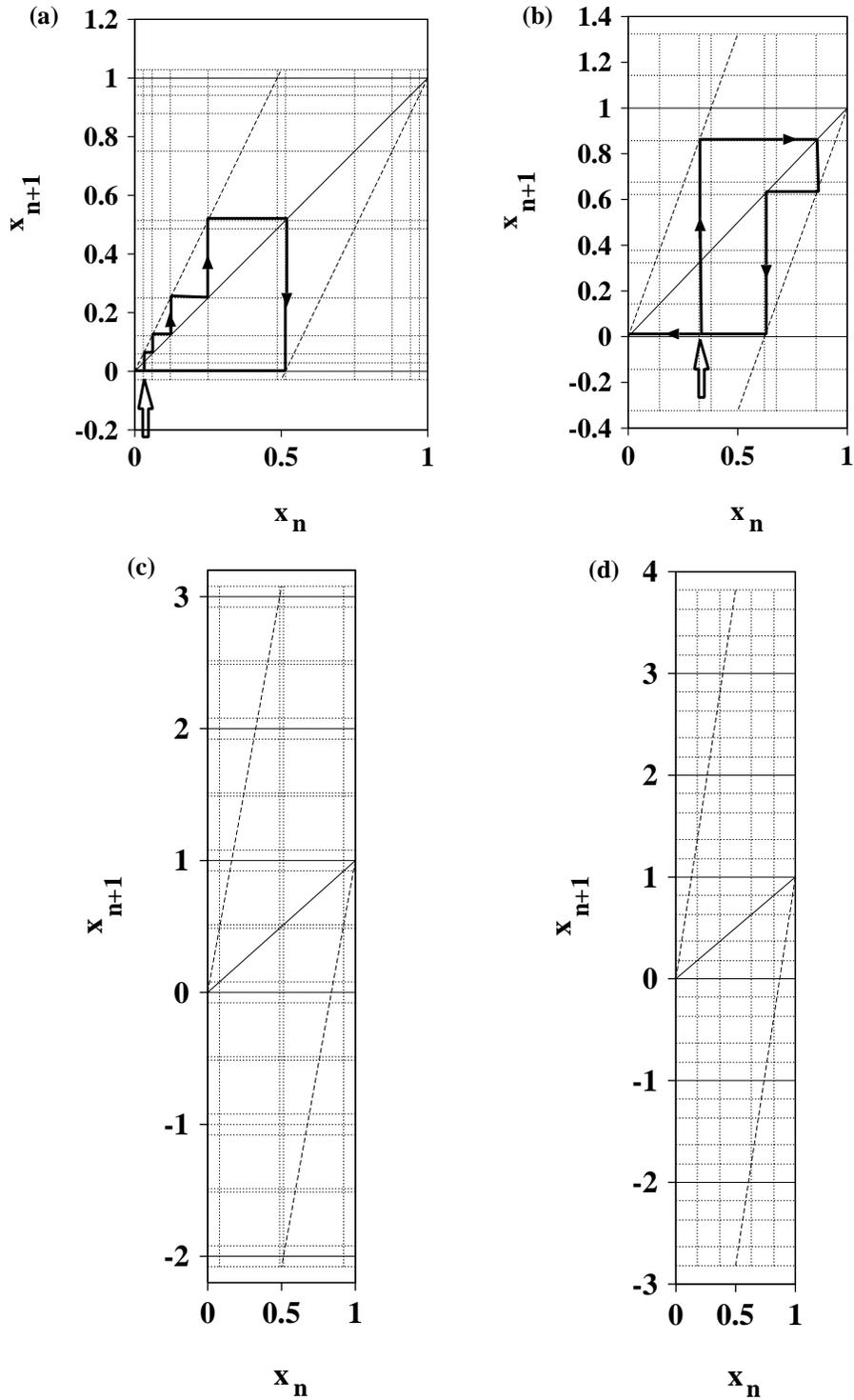}}
 
\vspace*{0.5cm}
\caption{\label{fig:mpex}Four examples of non-trivial Markov partitions of
map $\cal L$ at different values of the slope. Diagram (a) is for the
slope $a\simeq 2.057$, (b) for $a\simeq 2.648$, (c) for $a\simeq
6.158$ and (d) for $a=7.641$. In (a) and (b) the bold black lines with
the arrows show the generating orbits of the partitions, that is, the
orbits which define the single partition points (see text). The two
large arrows below the orbits indicate the respective initial
positions for the generating orbits.}
\end{figure}

In the previous section, the choice of simple partitions enabled the
construction of transition matrices. These matrices provided a way to
solve the Frobenius-Perron equation in a certain limit. However, so
far this method has only been applied to very special cases of map
$\cal L$, defined by integer slopes. This raises the question whether
an extension of this method to other values of the slope is
possible. For this purpose, the idea of choosing a suitable partition
of the map has to be generalized. Taking a look at Fig.\
\ref{fig:mp4} again, one observes that the graph of the map
``crosses'' or ``touches'' a vertical line of the grid only at some
grid points. Furthermore, the local extrema of the map, which are here
identical to the points of discontinuity, are situated on, or just
``touch'' horizontal lines of the grid, whereas other crossovers of
horizontal lines occur at no specific point. The same characteristics
can be verified, e.g., for the respective partition of slope $a=3$.
These conditions ensure that it is possible to obtain a correct
transition matrix from a partition, since to be modeled by a matrix, a
density of points, which covers parts of the partition completely, has
to get mapped in a way that its image again covers parts of the
partition completely, and not partially. This basic property of a
``suitable partition to construct transition matrices'' is already the
essence of what is known as a {\em Markov partition}:
\newtheorem{defs}{Definition}
\begin{defs}[Markov partition, verbal definition] 
For one-dimensional maps, a partition is a Markov partition if and
only if parts of the partition get mapped again {\rm onto} parts of
the partition, or {\rm onto} unions of parts of the partition
\cite{Beck}.
\label{def:mpv}
\end{defs}
A more formal definition of one-dimensional Markov partitions, as well
as further details, can be found in Refs.\ \cite{Bow79,Rue,dMvS}. The
next goal must be to find a general rule of how to construct Markov
partitions for map $\cal L$ at other, non-trivial parameter values of
the slope. Because of the periodicity of the chain of maps it suffices
to find a Markov partition for a single {\em box map}, i.e., for the
respective map in one box without applying the modulus to restrict it
onto the unit interval. Here, the fact can be used that the extrema,
which are the {\em critical points} of the box map, have to touch
horizontal lines, as explained before, which means that to obtain a
Markov partition the extrema have to get mapped onto partition points.
Since any box map of map $\cal L$ is symmetric with respect to the
point $(x,y)=(0.5,0.5)$, the problem reduces to considering only one
of the extrema in the following, e.g., the maximum. Changing the
height of the maximum corresponds to changing the slope of the
map. Therefore, if one needs to find Markov partitions for parameter
values of the slope, one can do it the other way around by the
following {\em Markov condition}:
\begin{defs}[Markov condition, verbal definition] \label{def:mcv}
For map $\cal L$, Markov partition values of the slo\-pe are
determined by choosing the slope such that the maximum of the box map
gets mapped onto a point of the partition again.
\end{defs}
In Fig.\ \ref{fig:mpex}, four examples of non-trivial Markov
partitions for map $\cal L$ are depicted with respect to their box
maps. One may check that the handwaving conditions which have been
extracted from the integer slope examples to motivate Definition
\ref{def:mpv} are fulfilled and especially that the partitions shown
in the figure obey the Markov partition definition
\ref{def:mpv}. The detailed structure of the partitions can be arbitrarily
complex. In Fig.\ \ref{fig:mpex} (a) and (b) a special orbit has been
marked by bold black lines with arrows. It represent what will be
called the {\em generating orbit} of a Markov partition: For the two
examples shown here, the starting point of this orbit is given by the
maximum of the preceding box map, since this maximum must also be a
partition point. The iterations of this orbit, as indicated by the
arrows in the figure, define the single partition points.  This
way, the number of partition parts is related to the number of
iterations of the generating orbit. In case of Fig.\
\ref{fig:mpex} (a) and (b) the orbit is eventually periodic, i.e., it
finally gets mapped onto the fixed point $x=0$, however, it can also
be periodic with a certain period, as, for example, in case of Fig.\
\ref{fig:mpex} (d) (period four). Thus, the generating
orbit is the key of how to find Markov partition values of the slope
in a systematic way.

On this basis, a general algebraic procedure to compute such values of
the slope can be developed. One starts with a further {\em topological
reduction} of the whole chain of boxes \cite{reduc}. Since map $\cal L$
is {\em old}, it is possible to construct the Markov partition for the
whole chain from a {\em reduced map}
\begin{equation}
\tilde{M}_a(\tilde{x}):=M_a(\tilde{x}) \quad \mbox{mod} \; 1 \label{eq:mod}
\end{equation}
via periodic continuation, where $\tilde{x} := x-[x]$ is the fractional
part of $x$, $\tilde{x}\:\in \:(0,1]$, and $[x]$ denotes the largest
integer less than $x$. Therefore, it remains to find Markov
partitions for map $\tilde{M}_a(\tilde{x})$ of the equation above. This can
be done in the following way: Let
\begin{equation}
\epsilon
:=min\left\{\tilde{M}_a(\frac{1}{2}),1-\tilde{M}_a(\frac{1}{2})\right\}\quad ,
\quad \epsilon \le \frac{1}{2}\quad ,\label{eq:epsdef}
\end{equation}
be the minimal distance of a maximum of the box map $M_a(\tilde{x})$ to
an integer value. With respect to the Markov condition given by
Definition \ref{def:mcv}, it is clear that $\epsilon$ has to be a
partition point. Since $\tilde{M}_a(\tilde{x})$ is point symmetric,
$1-\epsilon$ also has to be a partition point, and because of map
$\cal L$ being {\em old}, the fixed point $x=0$ is necessarily another
partition point. Thus, the reduced map governs its internal box
dynamics according to
\begin{equation}
\tilde{x}_{n+1}=\tilde{M}_a(\tilde{x}_n)\quad , \quad
\tilde{x}_n=\tilde{M}_a^n(\tilde{x}) \quad , \quad \tilde{x}\equiv \tilde{x}_0 \quad .
\end{equation}
Since $0,\epsilon$ and $1-\epsilon$ have to be partition points, the
Markov condition Definition \ref{def:mcv} can be formalized to
\begin{equation}
\tilde{M}_a^n(\epsilon)\equiv\,\delta\,\quad ,\quad
\delta\in\left\{0,\epsilon,1-\epsilon\right\} \quad , \label{eq:mcond}
\end{equation}
i.e., the generating orbit of a Markov partition is defined by the
initial condition $\epsilon$, its end point $\delta$ and the iteration
number $n$.  According to Eq.(\ref{eq:epsdef}), $\epsilon$ is
determined by the slope $a$. Therefore, for map $\cal L$ Markov
partition values of the slope can be computed as solutions of
Eq.(\ref{eq:mcond}). The evaluation of this equation can be performed
numerically as well as, to a certain degree, analytically. To obtain
analytical results for Markov partition values of the slope, one has
to determine the structure of the generating orbit in advance, i.e.,
one has to know whether it hits the left or the right branch of the
box map at the next iteration. Then one can write down an algebraic
equation which remains to be solved. For example, for the Markov
partition Fig.\ \ref{fig:mpex} (c), the generating orbit is determined
by
\begin{equation}
x_1=\tilde{M}_a(\epsilon)\quad , \quad \epsilon \le \frac{1}{2}\quad
\mbox{and} \quad 
\delta\equiv \tilde{M}_a(x_1)-3\quad , \quad x_1\le\frac{1}{2}
\end{equation}
with $a=2(3+\epsilon)$ and $\delta=0$ at iteration number $n=2$. This
leads to 
\begin{equation}
a^3-6a^2-6=0\quad , \quad a\ge 2\quad ,
\end{equation}
for which one may verify $a\simeq 6.158$ as the correct solution. This
way, all Markov partition values of the slope are the roots of
algebraic equations of $(n+1)$th order. More examples with analytical
solutions are discussed in App.\ \ref{app:dkmp}. Since one usually
faces the problem to solve algebraic equations of order greater than
3, numerical solutions of Eq.(\ref{eq:mcond}) are desirable, although
one should take into account that iterations of the reduced map
$\tilde{M}_a(\tilde{x})$ contain many discontinuities, due to the
original discontinuity of $\tilde{M}_a(\tilde{x})$ at $\tilde{x}=1/2$
as well as due to applying the modulus to $M_a(\tilde{x})$ in
Eq.(\ref{eq:mod}) \cite{rootf}.

With respect to
the three different end points $\delta$ of the generating orbit in the
formal Markov condition Eq.(\ref{eq:mcond}), three series of Markov
partitions can be distinguished. For each series one can increase the
iteration number $n$, and one can vary the range of the slope $a$
systematically. These three series have been used as the basis for
numerical calculations of the diffusion coefficient $D(a)$, as will be
explained below. However, there exist additional suitable end points
$\delta$ for the generating orbit. As an example, one can choose
$\delta$ to be a point on a two-periodic orbit,
\begin{equation}
\tilde{M}_a^2(\delta)=\delta\quad ,\quad a=2(1+\epsilon) \quad , \quad
0\le \epsilon \le \frac{1}{2} \quad \Rightarrow \quad 
\delta=\frac{1+2\epsilon}{4(1+\epsilon)^2-1}
\label{eq:2po}
\end{equation}
so that the generating orbit is again eventually periodic, but now
being mapped on a periodic orbit which is part of the Markov partition
instead of being mapped on a simple fixed point. This way, certain
periodic orbits can serve for defining an arbitrary number of new
Markov partition series with respect to the choice of respective new
end points $\delta$. On the other hand, the set of Markov partition
generating orbits is {\em not} equal to the set of {\em all} periodic
orbits. For example, for the range $2\le a \le3$, Eq.(\ref{eq:2po})
shows that there exists a two-periodic orbit for any slope $a$, but
{\em not any} maximum of the map in this range necessarily maps onto
this periodic orbit, as is already illustrated by Fig.\ \ref{fig:mpex}
(a) and (b), or by other simple solutions of Eq.(\ref{eq:mcond}),
respectively. This proves that Markov partition generating orbits are
in fact a subset of all periodic orbits of the map.

With respect to varying the iteration number $n$ and the end point
$\delta$, one can expect to get an infinite number of Markov partition
values of the slope. In fact, for certain classes of maps the
existence of Markov partitions can be considered as a natural property
of the map \cite{Sin68a,Bow1,Rue78,Rue}.
According to the explanations above, this does not seem to be
true for map $\cal L$. Instead, there is numerical evidence for the
following conjecture \cite{numev}:
\newtheorem{conj}{Conjecture}
\begin{conj}[Denseness property of Markov partitions] \label{conj:mpd}
For map $\cal L$, the Mar\-kov partition values of the
slope a are dense on the real line with $a\ge 2$.
\end{conj}
This denseness conjecture should ensure that it is possible to obtain
a representative curve for the parameter-dependent diffusion
coefficient $D(a)$ solely by computing diffusion coefficients at
Markov partition values of the slope. Conjecture \ref{conj:mpd} may
hold for all other class $\cal P$-maps as well. To do such
computations, one needs to construct the corresponding transition
matrices to the Markov partitions obtained, as it has been shown for
the slopes $a=3$ and $a=4$ of map $\cal L$. This can be done according
to the following rule: Take as an example any of the box map Markov
partitions illustrated in Fig.\ \ref{fig:mpex}, e.g., case (a). Any
dashed rectangle of this partition may be denoted as a {\em cell of
the partition}. These single cells correspond to the single entries,
or matrix elements, of the transition matrix to be obtained. The
transition matrix corresponding to this Markov partition can now be
constructed by checking where the graph of the map goes across a cell
of the partition, by counting the number of these occurrences in each
cell and by writing down these values as the matrix elements. For map
$\cal L$, usually these matrix elements will consist of zeroes and
ones, but the way they are defined here they can also take other
integer values, depending on the choice of the partition, as, e.g.,
illustrated in case of $a=4$, Eq.\ (\ref{eq:erf3}).

The construction of the box map transition matrix can be simplified by
taking the point symmetry of the box map into account. The transition
matrix of the full chain of chainlength $L$ again follows by periodic
continuation. These matrices can be denoted as {\em topological
transition matrices}, since they reflect purely the topology of the
map with respect to the Markov partitions, without involving any
transition probabilities at this point. In Refs.\
\cite{GH,Sin68a,Bow1,Rue78,Rue} mathematically rigorous definitions of
these transition matrices can be found. The property of map $\cal L$
being {\em old} induces a certain structure in the topological
transition matrices. They are said to be {\em banded square block
Toeplitz matrices}, i.e., they consist of certain submatrices, called
{\em blocks}, corresponding to the box map Markov partitions, and
these blocks are the same along diagonals of the topological
transition matrix parallel to the main diagonal, forming {\em bands}
\cite{Dav79,BeWa91,Tren85}. Applying periodic boundary conditions to
the chain of boxes defines a subclass of these Toeplitz matrices,
called {\em block circulants}, where each row is constructed by
cycling the previous row forward one block
\cite{BeKa52,Kow54,Dav79,BeWa91}, see, e.g., the matrices $T(4)$,
Eq.(\ref{eq:tm4}), as an example for a simple circulant and $T(3)$,
Eq.(\ref{eq:tm3}), for a block circulant. According to the transition
matrix method outlined in the previous section, it remains to solve
the eigenvalue problems of these matrices and to match the respective
eigenmodes to the ones of the diffusion equation for computing the
corresponding diffusion coefficients $D(a)$. Here, periodic boundary
conditions are of great advantage. Analytically, as mentioned before
and as shown in App.\ \ref{app:dkpbc}, there exists a general
procedure how to solve the eigenvalue problems of simple circulants
\cite{BeKa52,Kow54,Dav79}, and in some cases it is possible to reduce
the eigenvalue problem of a block circulant to that of a simple
circulant (see App.\ \ref{app:dkpbc} and \ref{app:dkmp}). If this
method works, it automatically yields ``nice eigenmodes'', i.e.,
eigenvectors of the form of sines and cosines with some fine
structure. These eigenmodes are similar to the eigenmodes of the
diffusion equation at this stage (see App.\ \ref{app:dkmp}), i.e.,
before iterating the matrices according to the Frobenius-Perron matrix
equation Eq.(\ref{eq:fpm}). The situation is quite different for
absorbing boundary conditions, where no such general procedure exists
(see App.\ \ref{app:dkabc}).

If analytical solutions of the eigenvalue problems are not possible
anymore, one can obtain numerical solutions. Well-known software
packages like NAG and IMSL provide subroutines to solve the eigenvalue
problems of these matrices. Unfortunately, the numerically obtained
results for the full spectra turned out not to be very reliable to a
certain extent: In comparison to analytical results for periodic
boundaries (see App.\ \ref{app:dkmp}), the NAG package does not
compute all eigenvectors correctly, i.e., in the numerical results
usually some linear independent eigenvectors are missing. Moreover,
both packages provide spectra of eigenvalues which, although partly
identical to the analytical solutions, differ in their full range
quantitatively to the ones calculated analytically, not taking any
degeneracy into account \cite{negev}. Such numerical problems seem to
be inherent to the class of non-normal Toeplitz matrices, as has
already been pointed out by Beam and Warming
\cite{BeWa91}. In fact, solving eigenvalue problems for Toeplitz
matrices can be considered as a field of active recent research in
numerical mathematics \cite{ReTr92,BaMo94}. However, solely for the
purpose of computing diffusion coefficients not the full spectra of
the transition matrices are required, but only the few largest
eigenvalues and eigenvectors are of interest. With respect to
eigenvectors, the IMSL package has been checked to be reliable in this
range, and with respect to eigenvalues, both packages provide exact
and identical numerical results, especially for the second largest
eigenvalue, which determines the diffusion coefficients. From a
practical point of view, the NAG package is considerably more
efficient in the use of computing time than IMSL and more flexible in
its subroutines. Therefore, for quantitative computations of
eigenvalues and diffusion coefficients NAG routines have been used,
whereas for eigenvector computations IMSL procedures have been
employed. For computations of diffusion coefficients, it is also
favorable to consider only the case of periodic boundary conditions,
i.e., solving eigenvalue problems for block circulants, respectively,
since it has already been discussed in Sect
\ref{txt:tmm} that absorbing boundaries lead to a poor convergence
rate of the diffusion coefficient with the chainlength $L$.

\begin{figure}
\epsfxsize=17cm
\centerline{\epsfbox{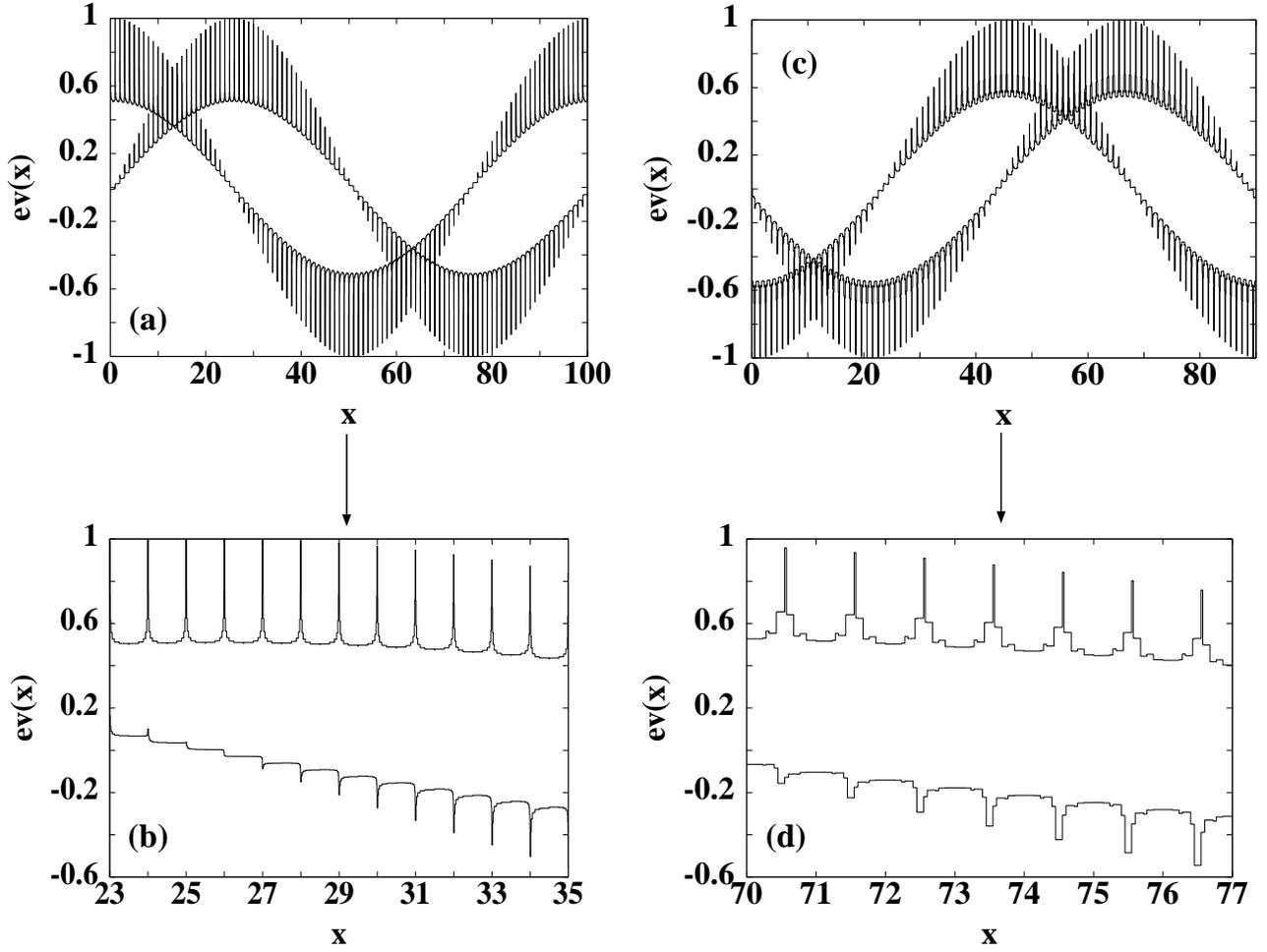}}
 
\vspace*{0.5cm}
\caption{\label{fig:emex}Second largest eigenmodes of map $\cal L$ at
two non-trivial Markov partition values of the slope with periodic
boundary conditions: full modes and magnifications of their fine
structures. For both parameter values there are two largest eigenmodes
which differ by a phase shift. Diagrams (a) and (b) are for slope
$a=3.0027$, chainlength $L=100$, (c) and (d) for $a=2.0148$,
chainlength $L=90$.}
\end{figure}

Fig.\ \ref{fig:emex} contains two examples of second largest
eigenmodes for chains of boxes with periodic boundaries and
non-trivial Markov partitions. Again, one gets ``nice'' second largest
eigenmodes, i.e., functions which behave like sines and cosines on a
large scale. However, the structure of these eigenmodes is much more
complex on a fine scale, as can be seen in the magnifications of
certain regions. The periodic continuation of the fine structure
suggests that it is related to the dynamics of the box map, and
therefore varies with changing the slope, whereas the general
large-scale behavior of the eigenmodes seems to be a property of the
chain of boxes which shows up independently from such microscopic
details. These characteristics have been checked numerically for a
variety of other Markov partition values of the slope and seem to be a
universal feature of map $\cal L$, and probably of all class $\cal
P$-maps. One may assume that the fine structure is somehow related to
the strength of the diffusion coefficient and that, on the other hand,
the universal large-scale structure of the eigenmodes is related to
the existence of diffusion coefficients for non-trivial Markov
partition values of the slope. In fact, the specific character of the
eigenmodes discussed above, which shows up in any analytical solution
of (block) circulants and which is supported by numerical results,
forms the basis for the following conjecture:
\begin{conj}[Existence of diffusion coefficients] \label{conj:edk}
Let $M_a(x)$ be a class $\cal P$-map. If for given
value of the slope the map is uniquely ergodic and if there
exists a Markov partition, then the map is diffusive.
\end{conj}
To our knowledge, so far no proof has been given in the literature for
the existence of diffusion coefficients in class $\cal P$-maps for
general value of the slope. However, dealing with a rigorous
foundation of the transition matrix method turns out to be intimately
connected to proving the existence of diffusion coefficients in this
class of dynamical systems. Without going into too much detail here,
some remarks are in order to provide at least a motivation for this
conjecture: The existence of Markov partitions guarantees that exact
transition matrices can be used. The restriction to class $\cal
P$-maps ensures that topological transition matrices can be
constructed in the simple way outlined before, and the {\em old}
property included in the definition of class $\cal P$ determines the
global structure of the topological transition matrices such that the
eigenmodes are ``nice'', at least for periodic boundary
conditions. The requirement to be uniquely ergodic establishes the
possibility of diffusion in the chain of boxes and confirms also the
uniqueness of the diffusion coefficient to be obtained (a simple
counterexample shows that not any chain of boxes with escape out of
one box is automatically diffusive).  Finally, the term diffusive
shall be understood in the sense that a diffusion coefficient exists
as defined by the statistical diffusion equation Eq.(\ref{eq:f2}),
which has been introduced to the dynamical system by successfully
performing the matching eigenmodes procedure outlined in the previous
section. Therefore, the main proposition of this conjecture is that
the matching eigenmodes procedure required by the first passage method
works for any value of the slope, if the respective conditions are
fulfilled. A corollary to this conjecture is that in the limit of time
$n$ and chainlength $L$ to infinity, the Frobenius-Perron equation of
the respective class $\cal P$-dynamical systems always provides
``nice'', i.e., the correct diffusive eigenmodes. As another
corollary, it follows that there is no anomalous diffusion in class
$\cal P$-maps, i.e., that normal diffusion is ``typical'' for such
piecewise linear maps. A rigorous mathematical proof of Conjecture
\ref{conj:edk} seems to be possible along the lines of first passage
and transition matrix method \cite{Kla95}.

Results based on this method shall be presented in the next
section. They have been verified by another numerical method
\cite{RKdiss,dcrc,Kla95}, by another analytical method which has been
implemented numerically \cite{RKdiss,tak}, as well as, to a certain
degree, by straightforward computer simulations \cite{RKdiss,Kla95}.
Meanwhile, the same results have been obtained by J.\ Groeneveld with
a different method \cite{Groe95}, and they have also partly been
reproduced by cycle expansion techniques \cite{Cvit95}.

\section[Results]{Fractal diffusion coefficients: results}\label{txt:fdk}
\begin{figure}
\epsfxsize=15cm
\centerline{\epsfbox{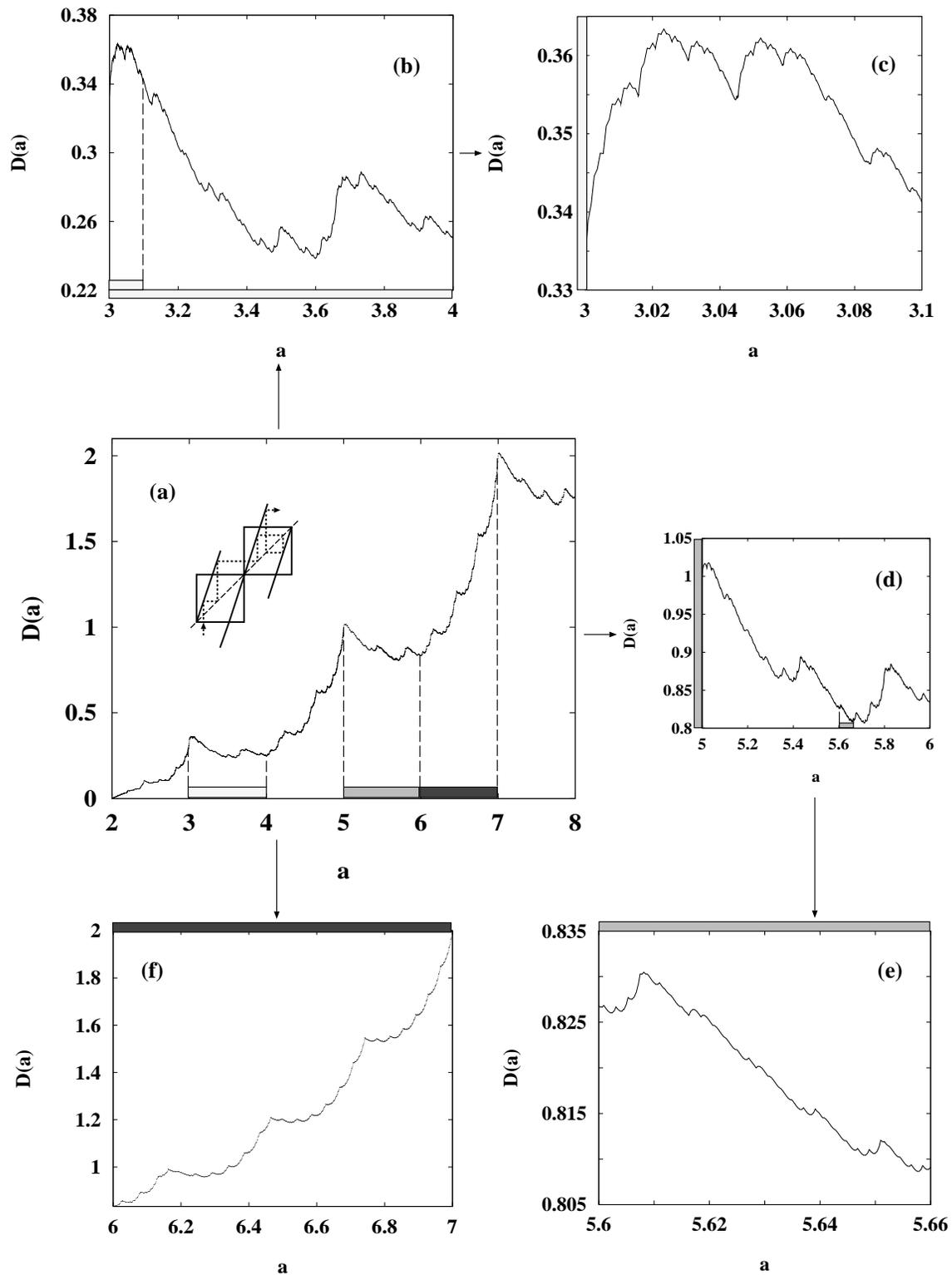}}
 
\vspace*{0.2cm}
\caption{\label{fig:dkmapl}
Parameter-dependent diffusion coefficient $D(a)$ of map $\cal L$ with
some blowups. In graph (b)-(e), the dots are connected with
lines. The number of data points is 7,908 for (a), 1,078 for (b), 476
for (c), 1,674 for (d), 530 for (e) and 1,375 for (f).}
\end{figure}

Based on the methods presented in the previous section, the
parameter-dependent diffusion coefficient $D(a)$ has been computed
numerically for map $\cal L$ for a broad range of values of the
slope. The main results are shown in Fig.\ \ref{fig:dkmapl}. The
numerical precision obtained depends on the convergence of the
diffusion coefficient with the chainlength, cf.\ Eq.(\ref{eq:dkl}),
and is better than $10^{-4}$ for each $D(a)$ so that error bars do not
appear in the diagrams. It should be emphasized that the numerical
method employed here was the first one by which these curves of $D(a)$
have been obtained. It is by far not the most efficient one of the
procedures developed up to now to compute deterministic diffusion
coefficients \cite{RKdiss,dcrc,Kla95,tak,Groe95}. However, it turns
out to be very useful as well to compute other deterministic transport
coefficients, e.g., chemical reaction rates \cite{GaKl}, where more
efficient methods fail.

Fig.\ \ref{fig:dkmapl} (a) shows the diffusion coefficient of map
$\cal L$ for values of the slope in the range $2\le a\le 8$.  The
strength of diffusion clearly increases globally by increasing the
slope from $a=2$ to $a=8$. This might be expected intuitively, since
the probability of a particle to escape out of a box, as well as the
mean distance a particle travels by performing a jump, are getting
larger by increasing the value of the slope
\cite{RKdiss,dcdd}. However, the increase of the diffusion coefficient
is not monotonic and consists of oscillations not only at integer
values of the slope, as has already been mentioned in Section
\ref{txt:fpm}, but also on much finer scales between integer
values. In fact, Fig.\ \ref{fig:dkmapl} (a) shows a certain regularity
in the appearance of ``wiggles'', i.e., local maxima and minima. If
one denotes the local maxima at odd integer slopes $a$ as wiggles of
$0$th order and any smaller local maxima systematically as wiggles of
higher order, one can find one maximum of first order below $a=3$,
three maxima of first order in the range $3\le a\le 5$, five maxima of
first order in the range $5\le a
\le 7$, \ldots . This regularity even persists to a certain extent on
finer scales, although according to a slightly different rule, as can
be seen, e.g., in the magnification Fig.\ \ref{fig:dkmapl} (f), $6\le
a\le 7$, where exactly six wiggles of second order appear between the
repsective wiggles of first order. In the same way, six wiggles of
third order can be observed in this region in further magnifications,
and similar structures show up in the region of $4\le a \le5$ with
four wiggles of second and four wiggles of third order. The region of
$2\le a \le 3$ is somewhat special and will be discussed
separately. Thus, while the number of wiggles of first order increases
by a step of two with increasing the slope, the number of wiggles of
higher order remains constant in the region between two respective
wiggles of first order, even by increasing the order of the wiggles to
be considered.  On the other hand, magnifications of other regions of
the slope show that the structure of the curve is not everywhere that
simple. For example, blow-ups of the regions $3\le a \le 4$, Fig.\
\ref{fig:dkmapl} (b), and $5 \le a \le 6$, Fig.\ \ref{fig:dkmapl} (d),
do not enable a clear distinction between ``wiggles of different
orders'' anymore. Instead, they provide more complex structures which
further magnifications, as, e.g., Figs.\ \ref{fig:dkmapl} (c) and (e),
reveal to be self-similar.

It can be summarized at this stage that different regions of the curve
exhibit different kinds of self-similarity, partly being fairly
simple, but partly also being highly non-trivial.  Thus, the results
of Fig.\ \ref{fig:dkmapl} suggest that the parameter-dependent
diffusion coefficient $D(a)$ for map $\cal L$ is fractal
\cite{Man82}. More evidence for the fractality of the curve can be
obtained in three different ways: Firstly, qualitative and
quantitative explanations for the wiggles in certain regions of the
slope will be provided, which ensure that these regions exhibit
non-trivial self-similar behavior.  This will be demonstrated in the
following. Secondly, it is striking to observe that especially
diagrams (c), (e), and (f) resemble graphs of some fractal functions,
which have been obtained in Refs.\ \cite{TG1,TG2,TA2} by working on
dynamical systems being very similar to the ones considered here. These
functions have been shown to possess fractal dimensions close to one
\cite{TA2}. On the one hand, this gives further evidence for the
fractality of the curves of Fig.\
\ref{fig:dkmapl}. On the other hand, this raises the new question
whether there is an analytical fractal representation for certain
regions of the curve, or maybe even for the full parameter-dependent
diffusion coefficient of map $\cal L$. This problem is discussed in
detail in Refs.\ \cite{RKdiss,tak}. Thirdly, it should be mentioned that
numerical computations of the box counting dimension (or capacity)
\cite{Ott} of the curve have been performed. The results
indicate that the curves shown in Fig.\ \ref{fig:dkmapl} (a) - (f)
have fractal dimensions $d$ very close to, but not equal to one in a
range of $d = 1+\Delta d\: , \: 0< \Delta d \le 10^{-2}$. Because of
the limited data set better values are difficult to get, especially
since the fractal dimension is expected to be close to one in this
case. Nevertheless, more detailed investigations of the fractal
dimension for various regions of the curve could be of much interest.
With respect to the magnifications in Fig.\
\ref{fig:dkmapl}, it may even be conjectured that the full
$D(a)$-curve has locally different values of fractal dimensions.

\begin{figure}
\epsfxsize=13cm
\centerline{\epsfbox{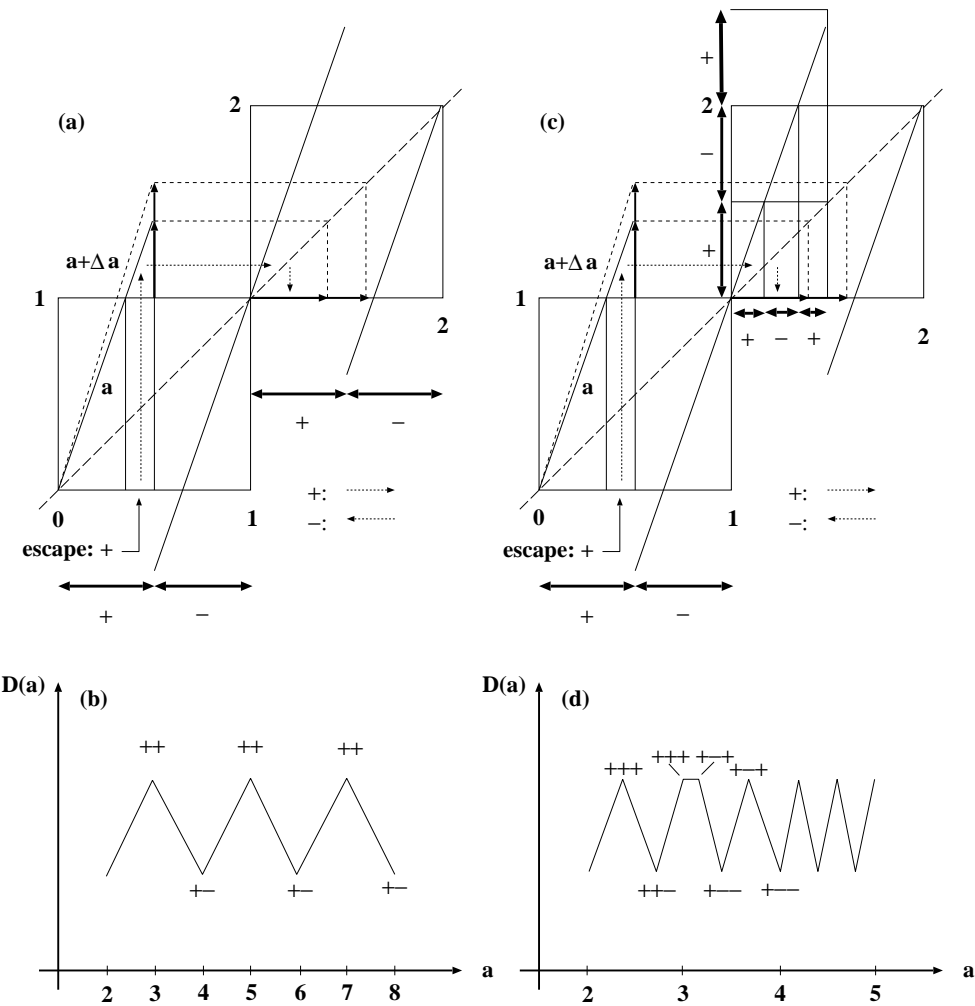}}
 
\vspace*{0.2cm}
\caption{\label{fig:plumil}First qualitative approach to understand
the structure of the parameter-dependent diffusion coefficient $D(a)$,
denoted as plus-minus method (see text). The variation of the
microscopic scattering process via changing the slope $a$ by $\Delta
a$ is heuristically related to variations in the strength of the
diffusion coefficient. The plus (minus) signs refer to subintervals
where forward (backward) scattering occurs at the next iteration of
the map (particle moves to the right, or left, respectively). The
qualitative argument is that the sequence of dominant forward or
backward scattering by varying the parameter induces oscillations in
the strength of the diffusion coefficient. Figs.\ (a) and (b) are for
one iteration, Figs.\ (c) and (d) for two.}
\end{figure}

Fig.\ \ref{fig:plumil} illustrates the principles of a first
qualitative approach to understand the occurrence of wiggles of $0$th
and $1$st order. It will be called {\em plus-minus approach}. The
basic idea of this approach is to establish a connection between the
appearance of wiggles in the $D(a)$-graphs and the occurrence of
certain dynamical correlations in the chain of boxes. These
correlations are a main feature of transport of particles from one box
to another, and they show up and disappear by varying the slope of the
map. In the following, particles will be referred to solely by their
positions, i.e., by points on the real line. Fig.\ \ref{fig:plumil}
(a) and (b) sketch correlations of $0$th order: As a starting point,
the escape of particles out of one box in one direction, i.e., to the
right, will be considered for varying the slope in the range $2\le a
\le 4$. Such an escape of points is related to a certain subinterval
of the box which will be called {\em escape region}, as is shown in
the figure.  If points get mapped to the right at the next iteration,
the respective subinterval will be denoted with a plus sign. The same
way, subintervals will be denoted with a minus if points get mapped to
the left. Therefore, the escape region marked in Fig.\
\ref{fig:plumil} is part of a plus region, and for small enough slope
after only one iteration points of it get mapped directly into another
plus region. This enhances diffusion, since particles can move
continuously in one direction, i.e., here to the right. The behavior
persists for increasing the slope up to $a=3$. For slopes above this
value, an increasing number of points of the escape region is now
mapped into the minus region of the next box. This way, one obtains a
``plus-minus'' correlation, which means that particles either get
slowed down, or even get scattered back into the previous box at the
next iteration, which is surely bad for obtaining a strong diffusion
coefficient. This game can be played by gradually increasing the value
of the slope and leads to the qualitative ``curve'' in Fig.\
\ref{fig:plumil} (b), which explains the oscillations at integer
slopes and the wiggles of $0$th order, respectively. The sequences
which mark the extrema in this graph give the symbolic dynamics of
orbits close to, but less than $x=1/2$ after one iteration with
respect to the reduced map Eq.(\ref{eq:mod}), where the region $0<x
\le 1/2$ has been labeled with a plus and $1/2<x\le 1$ with a minus.

In Fig.\ \ref{fig:plumil} (c) the number of iterations has been
increased to two. The method is the same as explained before, however,
a further distinction has been made after the first iteration: new
subintervals have been defined, which refer to points of the escape
region being mapped to another plus or minus region at the second
iteration. One can see that increasing the slope corresponds to
creating different plus-minus sequences for orbits close to, but less
than $x=1/2$. This leads to the particles being in a good or bad
position for going further in one direction with respect to the next
iteration, depending on the value of the slope. The $D(a)$-graph in
Fig.\
\ref{fig:plumil} (d) again gives the qualitative behavior of $D(a)$
to be expected with respect to the dynamical correlations after two
iterations, up to $a=5$. This result corresponds well to the number of
wiggles of first
order estimated in the respective regions of the slope. Again, the
plus-minus sequences give the symbolic
dynamics of points close to, but less than $x=1/2$ after two
iterations.

The plus-minus method works on this level as well for any higher
values of the slope and leads to a qualitative explanation of the
number of wiggles of first order for any region of the slope. To a
certain degree, it can even explain additional features of the
structure of the $D(a)$-curves: For example, in Fig.\ \ref{fig:dkmapl}
(b) one observes that the local maximum is not precisely at $a=3$,
although this could be expected from the results of the plus-minus
method of $0$th order. Actually, particles of the escape region close
to $x=1/2$ can still reach a good position for further movement in one
direction, even for slopes slightly above $a=3$. This is due to the
fact that, although such points get first scattered back into the
previous box after two iterations, here they are now in an excellent
position for further jumps to the right again. This way, these orbits
perform a kind of ``spiral'' and seem to be responsible for the
surprising fact that the odd integer slope values of $D(a)$ are {\em
not} precisely identical to the local extrema of the curve, but that
there is always a kind of overhang, i.e., a further increase of the
diffusion coefficient right above odd integer slopes, as, e.g., is
shown in detail in Fig.\ \ref{fig:dkmapl} (c).

Although the plus-minus method can be applied to achieve a qualitative
understanding of the wiggles of $0$th and $1$st order, further
refinements of this method to obtain wiggles of higher order generally
turned out not to be very promising. The main reason is that in case
of more iterations of points of the escape region, the dynamics is
getting quite complicated and is not easy to handle anymore in the
qualitative way illustrated in Fig.\
\ref{fig:plumil}. However, the basic idea of this method can be made
more quantitative by a procedure which shall be called {\em turnstile
dynamics}. The principle of turnstile dynamics is again to investigate
the appearance and disappearance of long-range dynamical correlations
by iterating points with respect to varying the slope. The new feature
is now that not the full intervals of all single boxes are taken into
account.  Instead, the analysis is restricted solely to the regions of
the boxes where transport of particles from one box to another occurs
in form of jumps. These regions are called {\em turnstiles}:
\begin{defs}[turnstile]
Turnstiles are the ``coupling regions'' of the single boxes of a
chain of class $\cal P$, where points of one unit interval get mapped
outside that particular interval into another unit interval.
\end{defs}
This notation has been adapted from the theory of transport of
two-dimensional twist maps, such as sawtooth maps, where turnstiles
are crucial for understanding large-scale diffusion
\cite{MMP84,CM89,Chen90,Meis92}. The escape region introduced above
in the context of the plus-minus method represents precisely one half
of such a turnstile. The main idea is to study the {\em
interaction} of turnstiles, i.e., by varying the slope it shall be
investigated whether one obtains ``good'' or ``bad'' conditions for
particles to get from one turnstile into another, or maybe even to get
mapped successively through a series of turnstiles. As before in case
of the plus-minus method, such dynamical correlations are expected to
show up in the curve for the parameter-dependent diffusion coefficient
$D(a)$. The advantage of turnstile dynamics is that it can be made
quantitative by exemplifying all turnstiles with certain points of
these regions. For instance, the peak of the turnstile one starts with
may be represented by the critical point, $x=1/2$. Now, one can try to
compute the slopes for which this point maps into other turnstiles
again, being exemplified by certain points, after certain numbers of
iterations.

\begin{figure}
\epsfysize=9cm
\centerline{\epsfbox{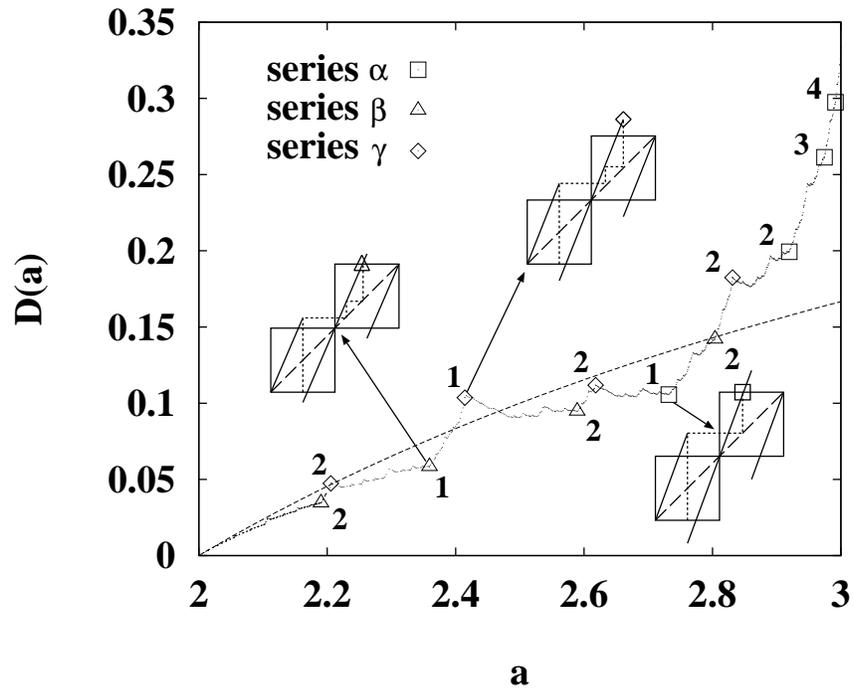}}
 
\vspace*{0.5cm}
\caption{\label{fig:tsdmapl}Blowup of the region of slope $a \leq
3$ for map $\cal L$ with the solution for a simple random walk model
(dashed line) and labels for parameter values which are significant
for ``turnstile dynamics'' (see text). Turnstile dynamics establishes
a quantitative relation between the local maxima and minima of the
parameter-dependent diffusion coefficient and the underlying
microscopic chaotic scattering process. For some parameter values,
the turnstile coupling is shown by pairs of boxes. The graph consists
of 979 single data points.}
\end{figure}

This has been done in detail for the region $2\le a \le 3$, as shown
in Fig.\ \ref{fig:tsdmapl}. The dashed line in the figure represents
the prediciton of $D(a)$ for a simple random-walk model as suggested
in Ref.\ \cite{SFK}, which is discussed in detail in Refs.\
\cite{RKdiss,dcrc,dcdd}. Note that, on a large scale, the model
correctly accounts for the behavior of $D(a)$ near $a=2$, but that
with respect to any fine structure, such a simple model is clearly
totally apart. One can see three distinct series of values of $a$ in
the figure.  To understand the nature of these series, one should
consider the orbit of the critical point. The first iterate of $x=1/2$
is in the second interval, $(1,2)$. The {\em series $\alpha$} values
of $a$ are defined by the condition that the second iterate of $x=1/2$
is at the leftmost point of the upward turnstile in the second
interval $(1,2)$ $(a=2.732)$, or that the third iterate is at the
corresponding point in the third interval $(a=2.920)$, etc.\ The
numbers on the $D(a)$-curve refer to the number of intervals the image
of $x=1/2$ has travelled before it gets to the appropriate point on
the turnstiles. {\em Series $\beta$} points are defined in a similar
way, but they are allowed to have two or more internal reflections
within an interval before reaching the left edge of a turnstile. {\em
Series $\gamma$} points are defined by the condition that some image
of $x=1/2$ has reached the rightmost edge of an upward turnstile,
i.e., some point $x=k+1/2$, where $k$ is an integer. One observes that
each series produces a cascade of apparently self-similar regions of
decreasing size, as the limits $a \rightarrow 2$ or $a \rightarrow 3$
are approached. These cascades provide a basis for a physical
understanding of the features of $D(a)$ in this region: Particles
leave a particular unit interval through a turnstile and undergo a
number of iterations before they are within another turnstile. Whether
they continue to move in the same or in the reverse direction at the
next and later turnstiles is a sensitive function of the slope of the
map. Thus, the fractal structure of the $D(a)$ curve is due to the
effects of long-range correlations among turnstiles, and these
correlations lead to changes of $D(a)$ on an infinitely fine scale.
We note that another way to understand this fractal structure is in
terms of so-called ``pruning'' of the microscopic deterministic
dynamics. That is, by varying the parameter $a$ certain types of
orbits may suddenly disappear. This means that with respect to a given
symbolic dynamics of the map certain symbol sequences, which identify
the orbits, do not exist anymore. This can be related to the
irregularities of the diffusion coefficient \cite{Cvit95,Tse94}.

One should note that series $\gamma$ points completely label the
maxima of higher order introduced before, and series $\beta$ points
mark the respective minima. This way, in the region of the slope $a\le
3$ the picture of quantitative turnstile dynamics is in full agreement
with the results obtained by the qualitative plus-minus method
outlined above (the agreement has been checked to persist at
least up to a level of extrema of second order). However, the
application of turnstile dynamics has its limits: First, this method
is of no use anymore for any higher value of the slope above
$a=3$. Thus, there is no other understanding of the structure in this
range than the one provided qualitatively by the plus-minus
approach. And second, even for values below $a=3$ turnstile dynamics
is quantitatively not completely correct: Apart from the lack of
explaining the existence of the overhang above $a=3$, a detailed
analysis reveals further ``tiny overhangs'' at maxima of higher order,
as, e.g., right above the maximum of first order in the region $2\le a
\le 3$ at $a=2.414$ \cite{tslim}.
In other words, the ``turnstile values'' marked in Fig.\
\ref{fig:tsdmapl} by series $\gamma$-points represent not the exact
local maxima of higher order of the curve. The true local maxima are
in fact shifted slightly to the right from these points, as in case of
$a=3$. The phenomenon of overhangs are further elucidated in Ref.\
\cite{RKdiss,tak}. However, apart from the qualitative remarks in the
context of the plus-minus approach and the additional insight provided
by the approach in Ref.\ \cite{RKdiss,tak}, a detailed explanation of
these overhang effects is still missing.

Since turnstile dynamics points seem to separate self-similar regions,
it is suggestive to use them as a tool to do some scaling. Series
$\alpha$-values are especially suitable for this purpose, because they
form a series of points which converges monotonically to $a=3$,
defining self-similar regions of decreasing size. These regions have
been scaled according to the size of the intervals of the slope and of
the respective diffusion coefficient intervals defined
by 
\begin{eqnarray}
\Delta a_i&:=&a(\mbox{series $\alpha$-point}\; (i+1))-a(\mbox{series
$\alpha$-point}\; i)\nonumber \\
\Delta D_i&:=&D(\mbox{series $\alpha$-point}\; (i+1))-D(\mbox{series
$\alpha$-point}\; i) \quad , \quad i=0,\ldots,4 
\end{eqnarray}
(note that $a=2$ has been taken here as series $\alpha$ point $0$). To
obtain scaling factors, the fractions
\begin{equation}
s_a(i):=\frac{\Delta a_{i-1}}{\Delta a_i} \quad \mbox{and}\quad
s_D(i):=\frac{\Delta D_{i-1}}{\Delta D_i} 
\end{equation}
have been computed. They led to the two series of values given in
Table \ref{tab:scal}. $s_a(i)$ seems to converge quite rapidly to a
value around $3$, whereas $s_D(i)$ approaches not that fast a value
apparently between $2$ and $3$, but this is no more than a guess based
on the first five values of two infinite series, since the data set of
slopes and $D(a)$-values is not sufficient for obtaining better
results. At least these values suggest that some quantitative scaling
is possible in this region.

At this point, it should be stressed that the region below $a=3$ is
special, compared to any other region of the slope: Firstly, the
structure of the curve is remarkably simple, as shown in Fig.\
\ref{fig:tsdmapl}. Secondly, the number of wiggles of higher order is
not constant with increasing order, but grows according to the
structure described by the turnstile dynamics performed above. This is
in contrast to the behavior of $D(a)$ in the ranges $4\le a \le 5$ and
$6\le a \le 7$, where one may have expected similar
generalities. Thirdly, the region below $a=3$ is the only one which is
simple enough such that turnstile dynamics can successfully be applied
at all, and this region seems to provide some simple scaling laws.
All this nice behavior suddenly breaks down at the value $a=3$, which
is marked by the largest overhang of the whole curve. Therefore, it
might be assumed that the point at $a=3$ separates regions of
fundamental different dynamical behavior of the map, i.e., the
dynamics seems to be sufficiently simple below, but suddenly gets
quite complicated above this value. In fact, there is further evidence
that such a transition exists, as is discussed in detail in Refs.\
\cite{RKdiss,dcdd}.

\begin{figure}
\epsfxsize=9.5cm
\centerline{\rotate[r]{\epsfbox{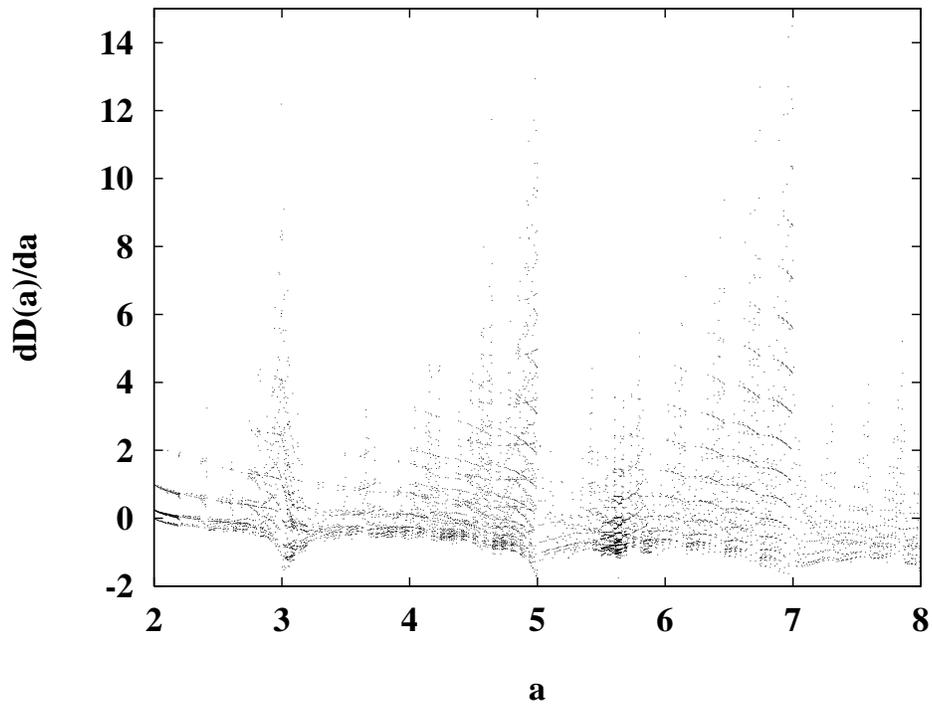}}}
 
\vspace*{0.6cm}
\caption{\label{fig:ddmapl}Numerical derivative of the
parameter-dependent diffusion coefficient $D(a)$ of map $\cal L$ with
respect to the slope $a$.}
\end{figure}

Finally, another interesting feature of the parameter-dependent
diffusion coefficient for map $\cal L$ shall be pointed out. To obtain
Fig.\ \ref{fig:ddmapl}, the first ``derivative'' of the $D(a)$-curve
of Fig.\ \ref{fig:dkmapl} has been computed with respect to the full
data set of $D(a)$-values available. This has been done in linear
approximation, i.e., two adjacent points of the data set have been
connected with straight lines, and the ``derivative'' $D'(a):=\Delta
D/\Delta a$ for $\Delta a \ll 1$ has been computed. However, since the
curve of $D(a)$ is expected to be nowhere differentiable with respect
to the slope, the derivative defined above should be better denoted as
a ``pseudo-derivative'', i.e., as a kind of mean value of $\Delta D$
over tiny regions of $\Delta a$ which are approaching zero. Thus, the
pseudo-derivative presented in Fig.\ \ref{fig:ddmapl} is
mathematically not well-defined. Nevertheless, the numerical results
turn out to be quite reasonable in the sense that they reflect to a
certain degree the fractal structure of the actual $D(a)$-curve. For
instance, the three ``bands'' right above $a=2$ seem to be due to the
``triangle-like'' self-similar structure of the region presented in
Fig.\ \ref{fig:ddmapl}, and the ``bursts'' at odd integer values
correspond to the occurrence of local maxima and to the deformations
of simple self-similar structures. We remark that in Fig.\
\ref{fig:ddmapl} the number of slopes for which $D(a)$-values have
been calculated is not homogeneously distributed over the whole range
$2\le a \le 8$, but that the number of points per interval is greater
in certain regions of the slope, as, e.g., right above $a=2$, $a=3$
and around $a=5.6$. These regions thus show up as slightly pronounced
parts in the derivative plot, however, the structure of the curve does
not seem to be affected by it. More details of this curve are
discussed in Refs.\ \cite{RKdiss,tak}.

\section{Summary}
\subsection{Conclusions}
(1) A simple model for deterministic diffusion has been discussed
where the microscopic scattering rules can be changed by varying a
single control parameter. The diffusion coefficient of this model has
been computed for a broad range of parameter values and shows a
fractal structure as a function of the slope of the map. This result
appears to be the first example of a dynamical system whose diffusion
coefficient has an unambiguously fractal structure.

(2) A general method to compute parameter-dependent diffusion
coefficients for a whole class of piecewise linear maps has been
developed. It is based on the first passage method, which provides the
definition of the diffusion coefficient for the dynamical system, in
combination with the use of Markov partitions and transition matrices,
which have been employed to solve the Frobenius-Perron equation of the
dynamical system. For periodic boundary conditions, the
parameter-dependent diffusion coefficient could be related to the
second largest eigenvalue of the topological transition matrix.  This
method provides analytical solutions in simple cases and is also
accessible to numerical implementations.

(3) The method described above has also been applied to absorbing
boundary conditions. Long-range boundary layers have been found in the
eigenmodes of the deterministic dynamical system. They also show up in
quantitative calculations of the diffusion coefficient.

(4) Certain limits of the first passage method in combination with the
use of transition matrices have been discussed: Drawbacks are
especially the restriction to certain initial probability densities
suitable for the application of transition matrices, as well as the
``external'' definition of the diffusion coefficient by the ``matching
eigenmodes'' procedure of first passage. It turned out that this
procedure is not well-defined anymore for smaller eigenmodes of the
dynamical system.

(5) A systematical way to find Markov partitions for the class of maps
under consideration has been developed. This method has been used as
the basis for computing the parameter-dependent diffusion coefficient
for the dynamical system mentioned above. For this map, as well as for
the whole class of maps under consideration, Markov partitions are
conjectured to be dense in the set of parameter values.

(6) A large number of eigenvalue problems of topological transition
matrices, based on Markov partitions, has been solved numerically to
compute the parameter-dependent diffusion coefficient for the
model. In the course of these calculations, the reliability of
well-known standard software routines for computing eigenvalue spectra
has been checked critically, and numerical uncertainties have been
pointed out.

(7) Certain large- and small-scale structures in the eigenmodes of the
topological transition matrices have been found. The large-scale
structures support the existence of statistical diffusion in the
dynamical system, whereas the small-scale structures refer to the
specific microscopic deterministic dynamics of the model system.
These results suggest that the strength of the fractal diffusion
coefficient is related to the fine-scale structure of the eigenmodes.

(8) A conjecture about the existence of diffusion coefficients for a
broad class of one-dimensional maps has been made. This conjecture may
shed more light on the origin of diffusion generated by a simple
deterministic dynamical system and may show a way how to put the
theory outlined in this article onto more solid mathematical grounds.

(9) Qualitative explanations for the structure of the
parameter-dependent diffusion coefficient over the full range of
parameter values have, to a certain extent, been provided by simple
heuristic considerations.

(10) A more refined ``turnstile dynamics'' has been developed as a
more quantitative approach to explain the structure of the
parameter-dependent diffusion coefficient. It works in certain regions
of the parameter values and provides a starting point for a scaling of
certain self-similar structures.

(11) By employing these qualitative and quantitative methods, certain
interesting features of the diffusion coefficient have been discussed,
i.e., the phenomenon of ``overhangs'' at local extrema, and the
special simple character of an ``initial region'' for small parameter
values, where diffusion sets in.

(12) The numerically computed pseudo-derivative of the
parameter-dependent diffusion coefficient seems to provide another
characteristic property of fractal diffusion coefficients.

\subsection{Outlook}
The class of one-dimensional piecewise linear maps we have studied
here by analyzing an example appears to be the most simple type of
deterministic diffusive systems one can think of. Nevertheless, we
have shown that the diffusion coefficient of such a map changes in a
fractal way by varying a control parameter. Starting from this
fundamental result, there are at least two directions in which our
research can be pursued: One way is to study whether other transport
quantities, like electric conductivities, chemical reaction rates, or
magnetoresistancies can exhibit such an irregular behavior as
well. Another way is to investigate whether fractal transport
coefficients exist in more complicated, and thus more realistic,
dynamical systems.

First steps in these directions have already been taken: For example,
a bias has been added to the simple map discussed in this paper. This
generates an average current of particles which again exhibits fractal
structures by varying the bias as a parameter
\cite{Kla95,Groe95}. Moreover, for small enough bias the current can
run opposite to the bias \cite{Kla95,Groe95}, and for other parameter
values the diffusion coefficient is zero with nonzero current
\cite{Kla95}. Deterministic diffusion coefficients and deterministic
currents which change irregularly by varying respective parameters
have also been found in parameter-dependent two-dimensional multibaker
models \cite{GaKl,KlTe97}. In their transport properties, these models
are closely related to the class of one-dimensional maps discussed
here. However, in addition they provide more physical features like
being time-reversible \cite{KlTe97,Kla95} and being area-preserving
or dissipative in a well-defined sense. Negative currents have been
observed in these systems as well, and a parameter-dependent
diffusive-reactive multibaker yields chemical reaction rates which are
also fractal in a parameter \cite{GaKl}.

In the periodic Lorentz gas, however, there is up to now no clear
indication about a fractal behavior of transport coefficients. First
results of computer simulations for the diffusion coefficient with
respect to varying the density of scatterers show a very smooth curve,
which indicates that if there are fractal fluctuations in the
parameter at all they must occur on a very fine scale
\cite{DeKl96}. On the other hand, for the thermostatted periodic
Lorentz gas with an electric field computer simulations of several
groups show clearly a very irregular behavior of the conductivity by
varying the field strength \cite{MH87,LNRM95,DeGP95,DeMo97}. In one
case, the numerical results could even be confirmed by calculations
based on cycle expansions \cite{DeMo97}.

Whether deterministic phenomena of this kind play a role in real
statistical, experimentally accessible systems is a very open
question. Following the chaotic hypothesis of Gallavotti and Cohen
\cite{GaCo95a,GaCo95b}, one may believe that these phenomena are rather
due to the simplicity of the models and should eventually disappear if
the systems are getting more complex. Respectively, we would expect
that certain necessary conditions must be fulfilled for systems to
exhibit characteristics of fractal transport coefficients which may
contradict the spirit of the chaotic hypothesis, such as being
spatially periodic, low-dimensional, and such that particle-particle
interactions are not of main importance. Physical systems of this kind
could - to a certain extent - already be realized experimentally in
form of so-called antidot lattices. Here, magnetoresistancies which
fluctuate irregularly by varying the field strength have already been
observed experimentally in a classical limit \cite{Weis91,WLR97}, and
to a certain extent they have been explained theoretically by
identifying special orbits in the microscopic dynamics
\cite{Geis90,FlSS}. Another candidate of a system where certain
irregularities in transport could be of a deterministic origin are
so-called ratchets, where negative currents have already been found
experimentally, as well as in theoretical models (see, e.g., Refs.\
\cite{HaBa96,JuHa96} and further references therein). In fact, it can
be argued that there exists a relation between certain types of
ratchets and the class of one-dimensional maps (supplemented by a
bias) studied here \cite{Kla95,PJ}.

Fractal transport coefficients in one-dimensional maps actually appear
to be stable with respect to imposing different kinds of random
perturbations on the system \cite{Kla95}. This means that the fractal
structure gradually smoothes out by increasing the perturbation
strength and thus survives in form of irregular oscillations on finer
scales if the perturbation is small enough. Although there are
exceptions to this behavior \cite{Rado96}, we believe this to be the
typical scenario of how a possible fractality of parameter-dependent
transport coefficients may appear if the system is not completely
deterministic.\\

\section*{Acknowledgments} 
This work is part of a Ph.D.\ thesis performed at the Technical
University of Berlin \cite{RKdiss}. R.K.\ and J.R.D.\ want to thank
Chr.\ Beck, L.\ Bunimovich, P.\ Gaspard, C.\ Grebogi, B.\ Hunt, H.E.\
Nusse, R.\ Kapral, A.\ Pikovsky, and J.\ Yorke for helpful
discussions. Special thanks go to J.\ Groeneveld for enlightening
e-mail discussions and for critical comments on parts of this
work. R.K.\ gratefully acknowledges financial support by the German
Academic Exchange Service (DAAD), by the NaF\"{o}G commission Berlin,
and by the Deutsche Forschungsgemeinschaft (DFG). He also wants to
thank the Institute for Physical Science and Technology (IPST),
College Park, and T.R. Kirkpatrick for their hospitality during a 1
1/2-year stay when this work got started, and for support in several
subsequent visits. Finally, R.K.\ wants to thank S.\ Hess and E.\
Sch\"oll for their interest in this work and for their continuous
support.

\begin{appendix}
\section[Transition matrix method]{Transition matrix method for
calculating diffusion coefficients}\label{app:dkbc}
In this Appendix, we deal with the problem of how to solve
analytically the eigenvalue problems of the topological transition
matrices $T(a)$ formally introduced in Section
\ref{txt:tmm}. These transition matrices are the key ingredients of
the Frobenius-Perron matrix equation, Eq.\ (\ref{eq:fpm}), and their
eigenmodes and eigenvalues determine the deterministic diffusive
dynamics of the map at the respective parameter value of the slope, as
outlined in Section \ref{txt:fpm}. Two simple examples of such
transition matrices have already been given by Eq.\ (\ref{eq:tm4}) and
Eq.\ (\ref{eq:tm3}), the problem of transition matrices for more
complicated Markov partitions has been discussed in Section
\ref{txt:mp}.

In the first subsection of this Appendix, we will consider integer
slopes with periodic boundary conditions. In this case, trivial Markov
partitions can be found for which the corresponding transition
matrices have a simple structure. We analytically solve the eigenvalue
problems of these transition matrices by standard methods and compute
the parameter-dependent diffusion coefficients as defined by the first
passage method. In the second subsection, we solve the eigenvalue
problems of the respective transition matrices for two integer slopes
with absorbing boundary conditions, namely for $a=3$ and $a=4$. In the
third subsection, we discuss the problem of non-trivial Markov
partitions for non-integer slopes. We depict Markov partitions for
some special values of the slope, sketch the corresponding transition
matrices, outline how to solve their eigenvalue problems analytically,
and give the results for their eigenvalues and diffusion coefficients.

\subsection{Integer slopes with periodic boundary
conditions}\label{app:dkpbc}
As discussed in Section \ref{txt:sfp}, for periodic boundary
conditions all topological transition matrices corresponding to Markov
partitions are block circulants. For certain values of the slope it is
possible to reduce these block circulants to simple circulants for
which it is known how to solve their eigenvalue problems
\cite{BeKa52,Kow54,Dav79}. We employ here the approach of Berlin and
Kac as described in App.\ A of Ref.\ 5\cite{BeKa52}. We summarize
their main formulas and use them first to solve the problem of even
integer slopes for periodic boundary conditions. We then treat
analogously the slightly more complicated case of odd integer slopes.

Let $T$ be a cyclic matrix of the type
\begin{equation}
T:=
\left( \begin{array}{cccccc} t_1 & t_2 & t_3 & \cdots  &t_{L-1} & t_L \\
t_L & t_1 & t_2 & \cdots & t_{L-2} & t_{L-1} \\
t_{L-1} & t_L & t_1 & \cdots & t_{L-3} & t_{L-2} \\
\vdots & & \vdots & & \vdots & \vdots \\
t_2 & t_3 & t_4 & \cdots & t_L & t_1  \\ 
\end{array} \right) \quad . \label{eq:ctm}
\end{equation}
We want to calculate the eigenvalues $\chi_m$ and the eigenvectors
$|\phi_m>$ associated with $T$, that is,
\begin{equation}
T\,|\phi_m>=\chi_m\,|\phi_m> \quad , \quad m=0,\ldots,L-1 \quad
. \label{eq:evpgt} 
\end{equation}
According to Ref.\ \cite{BeKa52}, the eigenvalues are given by
\begin{equation}
\chi_m=\sum_{s=1}^L t_s r_m^{s-1} \label{eq:evgt}
\end{equation}
with
\begin{equation}
r_m=\exp\left(i\frac{2\pi m}{L}\right) \quad , \label{eq:rmgt}
\end{equation}
and the corresponding eigenvectors are
\begin{eqnarray}
|\phi_m>&=&(\phi_m^1,\phi_m^2,\ldots,\phi_m^k,\ldots,\phi_m^L)^*\quad
, \quad \phi_m^k=\tilde{a}_m\phi_{m,1}^k+\tilde{b}_{m}\phi^k_{m,2}\quad ,
\nonumber \\  
& & \phi_{m,1}^k:=\cos(\theta_m(k-1))\quad ,\quad
\phi_{m,2}^k:=\sin(\theta_m(k-1))\quad ,\quad \nonumber \\
& & k=1,\ldots,L  \quad , \quad \theta_m:=\frac{2\pi m}{L} \label{eq:emgt}
\end{eqnarray}
with $\tilde{a}_m$ and $\tilde{b}_m$ to be fixed by suitable normalization
conditions.

We now compute the diffusion coefficient for all even integer slopes
$a=2k\: , \: k\in N$, of map $\cal L$ by solving the eigenvalue
problem of the respective general transition matrix $T(a)$ of the
system. This transition matrix can be constructed as discussed in
Section \ref{txt:tmm} by using the same Markov partition for all
slopes, that is, the one which is depicted in Fig.\ \ref{fig:mp4}. We
find that the matrix corresponding to this Markov partition is a
simple circulant with matrix elements of
\begin{equation}
t_1=2 \: , \: t_2=1 \: , \: \ldots \: , \: t_{a/2}=1 \: , \:
t_{a/2+1}=0 \: , \: \ldots \: , \: t_{L-a/2+1}=0 \: , \: t_{L-a/2+2}=1
\: ,  \: \ldots \: , \: t_L=1 \quad .
\end{equation}
The $(s+a/2-1)$th row, $1\le s\le L-a/2$, of the corresponding
eigenvalue problem defined by Eq.\ (\ref{eq:evpgt}) is thus determined
by
\begin{equation}
\phi_m^s+\ldots+\phi_m^{s+a/2-2}+2\phi_m^{s+a/2-1}+\phi_m^{s+a/2}+
\ldots+\phi_m^{s+a-2}=\chi_m\phi_m^{s+a/2-1} \quad .
\end{equation}
According to Eq.\ (\ref{eq:evgt}), the eigenvalues are 
\begin{equation}
\chi_m=2+r_m+\ldots+r_m^{(a-2)/2}+r_m^{L-(a-2)/2}+\ldots+r_m^{L-1}
\quad , 
\end{equation}
and by using Eq.\ (\ref{eq:rmgt}) we obtain explicitly
\begin{equation}
\chi_m=2+2\sum_{s=1}^{(a-2)/2}\cos (\theta_m s) \simeq
a-\theta_m^2\frac{a(a-1)(a-2)}{24} \quad (L\to\infty) \quad . 
\end{equation}
The corresponding eigenvectors are given by Eq.\ (\ref{eq:emgt}).
Note that the largest eigenvalue $\chi_0$ is equal to the slope of the
map. This is a consequence of the fact that the topological transition
matrices discussed here can be mapped onto stochastic transition
matrices and that the Perron-Frobenius theorem for non-negative
matrices applies \cite{PG1,Gant71}. It can be proven that this
property holds for any topological transition matrix of map $\cal L$
which is defined on the basis of Markov partitions with periodic
boundary conditions \cite{Kla95}. According to the matrix
Frobenius-Perron equation, Eq.\ (\ref{eq:fpm}), the corresponding
largest eigenmode determines the equilibrium state of the system,
which is here simply uniform. In analogy to Eq.\ (\ref{eq:dec4}), the
second largest eigenvalue gives the decay rate, and the respective
second largest eigenmode governs the diffusive transport in the
map. For slope $a=2k$ we thus obtain a decay rate of
\begin{equation}
\gamma_{dec}(a)=\ln\frac{a}{\chi_{1}(a)}\simeq
\frac{4\pi^2}{L^2}\frac{(a-1)(a-2)}{24} \quad (L\to\infty) \quad .  
\end{equation}
With Eq.\ (9), this gives a diffusion coeficient of
\begin{equation}
D(a)=\frac{L^2}{4\pi^2}\gamma_{dec}(a)=\frac{(a-1)(a-2)}{24} \quad .
\end{equation}
We now want to do the same calculation for all odd integer slopes
$a=2k-1 \: , \: k\in N$, again by employing periodic boundary
conditions. As a Markov partition for all these slopes we use the same
partitioning underlying the transition matrix as given by Eq.\
(\ref{eq:tm3}), that is, its parts are all of length 1/2. Here, the
corresponding $a$-dependent transition matrix is a block circulant
where the single blocks consist of 2x2-matrices. To illustrate the
general structure of this matrix, we give the first rows and columns
of the special case $a=5$ which is
\begin{equation}
T(5)=
\left( \begin{array}{ccccccccccc} 
1 & 1 & 0 & 1 & 0 & 0 & 0 & 0 & 0 & 0 & \cdots \\
1 & 1 & 0 & 1 & 0 & 1 & 0 & 0 & 0 & 0 & \cdots \\
1 & 0 & 1 & 1 & 0 & 1 & 0 & 0 & 0 & 0 & \cdots \\
1 & 0 & 1 & 1 & 0 & 1 & 0 & 1 & 0 & 0 & \cdots \\
1 & 0 & 1 & 0 & 1 & 1 & 0 & 1 & 0 & 0 & \cdots \\
0 & 0 & 1 & 0 & 1 & 1 & 0 & 1 & 0 & 1 & \cdots \\
0 & 0 & 1 & 0 & 1 & 0 & 1 & 1 & 0 & 1 & \cdots \\
\vdots & & & & \vdots & & & & \vdots \\
\end{array} \right) \quad . \label{eq:tm5}
\end{equation}
We now write down the eigenvalue equation, Eq.\ (\ref{eq:evpgt}), for
this matrix. By using the notation $u_m^k$ for the $k$th odd
component of $|\phi_m>$ and $v_m^k$ for the $k$th even component we
obtain for $k$ odd, $1\le k \le L-3\: , \: 0\le m\le L-1$,
\begin{eqnarray}
u_m^k+u_m^{k+1}+v_m^{k+1}+v_m^{k+2}+v_m^{k+3}&=&\chi_m v_m^{k+1}
\nonumber \\
u_m^k+u_m^{k+1}+u_m^{k+2}+v_m^{k+2}+v_m^{k+3}&=&\chi_m u_m^{k+2}\quad ,
\end{eqnarray}
supplemented by periodic boundary conditions for the respective first
and last rows of the matrix. This leads to $v_m^k=u_m^{k+1}$, yielding
an equation for $u_m^k$ which reads
\begin{equation}
u_m^k+u_m^{k+1}+u_m^{k+2}+u_m^{k+3}+u_m^{k+4}=\chi_m u_m^{k+2}\quad ,
\label{eq:tm5s} 
\end{equation}
again by providing respective periodic boundary conditions. Thus, we
have reduced the eigenvalue problem for the initial block circulant of
Eq.\ (\ref{eq:tm5}) to the eigenvalue problem of a simple circulant as
given by Eq.\ (\ref{eq:tm5s}). The same reduction procedure can be
carried out for general odd integer value of the slope. The reduced
eigenvalue equation then reads
\begin{equation}
u_m^k+\ldots +u_m^{k+a-1}=\chi_m u_m^{k+(a-1)/2} \quad ,
\end{equation}
supplemented by respective periodic boundary conditions. With the
Berlin-Kac method we obtain for the eigenvalues
\begin{equation}
\chi_m=1+2\sum_{s=1}^{(a-1)/2}\cos(\theta_m s) \simeq a-\frac{4\pi^2
m^2}{L^2}\frac{a(a^2-1)}{24} \quad (L\to\infty)\quad , \quad 0\le m\le
L-1 \quad . \label{eq:oddev}
\end{equation}
The eigenvectors are given here by 
\begin{eqnarray}
|\phi_m>&=&(u_m^1,v_m^1,\ldots,u_m^k,v_m^k,\ldots,u_m^L,v_m^L)^*\quad
, \quad u_m^k=\tilde{a}_m u_{m,1}^k+\tilde{b}_{m} u^k_{m,2}\quad , \quad
v_m^k=u_m^{k+1} \quad , \nonumber \\  
& & u_{m,1}^k:=\cos(\theta_m(k-1))\quad ,\quad
u_{m,2}^k:=\sin(\theta_m(k-1))\quad , \nonumber \\ 
& & k=1,\ldots,L  \quad , \quad \theta_m:=\frac{2\pi}{L}m 
\end{eqnarray}
with $\tilde{a}_m$ and $\tilde{b}_m$ to be fixed by suitable normalization
conditions. For the decay rate we get
\begin{equation}
\gamma_{dec}(a)\simeq \frac{4\pi^2}{L^2}\frac{(a^2-1)}{24} \quad
(L\to\infty) \quad ,
\end{equation}
which leads to a diffusion coeficient of
\begin{equation}
D(a)=\frac{a^2-1}{24} \quad .
\end{equation}

\subsection{Integer slopes with absorbing boundary
conditions}\label{app:dkabc} 

For absorbing boundary conditions, the corresponding transition
matrices are no block circulants anymore, but they belong to the
broader class of banded square block Toeplitz matrices, as pointed out
in Section \ref{txt:tmm}. For these matrices no general methods are
known for solving their eigenvalue problems analytically. However, in
certain cases analytical solutions can still be obtained by
straightforward calculations, as we will show for the two integer
values of slope $a=3$ and $a=4$.

We first consider the case $a=4$. The transition matrix for this
parameter value is identical to the one given by Eq.\ (\ref{eq:tm4})
except that the upper right and the lower left corners are filled with
zeroes because of absorbing boundaries. The eigenvalue problem of this
matrix can now be solved in analogy to the calculations performed by
Gaspard in Ref.\ \cite{PG1}. The eigenvalue equation, Eq.\
(\ref{eq:evpgt}), reads here
\begin{equation}
\phi_m^k+2\phi_m^{k+1}+\phi_m^{k+2}=\chi_m \phi_m^{k+1} \quad , \quad
0\le k\le L-1 \quad , \label{eq:phabs}
\end{equation}
supplemented by the absorbing boundary conditions
$\phi_m^0=\phi_m^{L+1}=0$. Since this equation is of the form of a
discretized ordinary differential equation of degree two we make the
ansatz
\begin{equation}
\phi_m^k=a \cos (k\theta) + b \sin (k\theta) \quad , \quad 0\le k \le
L+1 \quad . \label{eq:tmans}
\end{equation}
The two boundary conditions then lead to
\begin{equation}
a=0 \quad \mbox{and} \quad \sin ((L+1)\theta)=0
\end{equation}
yielding 
\begin{equation} 
\theta_m=\frac{m \pi}{L+1} \quad , \quad 1\le m\le L \quad .
\end{equation}
The eigenvectors are then determined by
\begin{equation}
\phi_m^k=b \sin(k\theta_m) \quad  
\end{equation}
with $b$ as the normalization constant. Putting this equation into Eq.\
(\ref{eq:phabs}) gives as the eigenvalues
\begin{equation}
\chi_m=2+2\cos \theta_m \quad .
\end{equation}
Note that in case of absorbing boundary conditions the largest
eigenvalue is not equal to the slope of the map, but determines the
escape rate of the system. Correspondingly, the largest eigenmode is
the diffusive mode of the map. However, for $L\to\infty$ the largest
eigenvalue goes to the exact value of the slope, which therefore
serves as an upper limit of the eigenvalue spectrum. This is
conjectured to be true for any topological transition matrix of map
$\cal L$ which is defined on the basis of Markov partitions by
employing absorbing boundary conditions \cite{Kla95}. In the limit of
chainlength $L\to\infty$ these results lead to the escape rate and
diffusion coefficient presented in Section
\ref{txt:tmm}, Eqs.\ (\ref{eq:escabs} -- \ref{eq:dkl}).

We now treat analogously the case of slope $a=3$ for absorbing
boundary conditions. We know from the previous subsection that for odd
integer slopes the Markov partitions, and thus the respective
transition matrices, are a bit more complicated. The transition matrix
for $a=3$ is identical to the one given by Eq.\ (\ref{eq:tm3}) except
that the upper right and lower left corners are filled with zeroes
because of absorbing boundaries, as before.

To write down the eigenvalue equation, Eq.\ (\ref{eq:evpgt}), for this
matrix we use the same notation as in the previous subsection for odd
integer slopes. With $u_m^k$ being the $k$th odd component of
$|\phi_m>$ and $v_m^k$ being the $k$th even component we obtain for
$k$ odd, $0\le k \le L$, 
\begin{eqnarray}
u_m^k+v_m^k+v_m^{k+1}&=&\chi_m v_m^k \nonumber \\
u_m^k+u_m^{k+1}+v_m^{k+1}&=&\chi_m u_m^{k+1}\quad ,
\end{eqnarray}
supplemented by the absorbing boundary conditions
$u_m^0=v_m^{L+1}=0$. This again leads to $v_m^k=u_m^{k+1}$, yielding
an equation for $u_m^k$ which reads
\begin{equation}
u_m^k+u_m^{k+1}+u_m^{k+2}=\chi_m u_m^{k+1}\quad , \quad 0\le k\le L
\quad , \label{eq:evabs3}
\end{equation}
with the respective absorbing boundary conditions. We again use Eq.\
(\ref{eq:tmans}) as an ansatz to solve this equation. The two
boundary conditions then lead to
\begin{equation}
a=0 \quad \mbox{and} \quad \sin ((L+2)\theta)=0
\end{equation}
yielding 
\begin{equation}
\theta_m=\frac{m\pi}{L+2}\quad , \quad 1\le m\le L+1 \quad .
\end{equation}
The eigenvectors are then determined by
\begin{eqnarray}
u_m^k&=&b \sin(k\theta_m) \nonumber \\
v_m^k&=&b \sin((k+1)\theta_m)
\end{eqnarray}
with $b$ as the normalization constant. Putting this equation into Eq.\
(\ref{eq:evabs3}) gives as eigenvalues
\begin{equation}
\chi_m=1+2\cos \theta_m \quad .
\end{equation}
In the limit of chainlength $L\to\infty$ this leads to the results for
escape rate and diffusion coefficient presented in Section
\ref{txt:tmm}, Eqs.\ (\ref{eq:escabs} -- \ref{eq:dkl}).

We remark that we do not have analytical solutions of the eigenvalue
problems for integer values of the slope above $a=4$ with absorbing
boundary conditions. Here, the ansatz of Eq.\ (\ref{eq:tmans}) does
not seem to be sufficient because of long-range boundary layers which
are induced by the absorbing boundary conditions, see also the remarks
in Section \ref{txt:tmm} to this problem.

\subsection[Markov partitions]{Non-trivial Markov partitions with
periodic boundary conditions}\label{app:dkmp}

In this subsection, we discuss some examples of non-trivial Markov
partitions where analytical solutions of the respective eigenvalue
problems of the transition matrices can still be obtained
analytically. This can be done by reducing block circulants onto
simple circulants in the way illustrated in the previous subsections
for odd integer slopes. In the following, we will only consider
periodic boundary conditions, because then the general method of
Berlin and Kac to solve the respective eigenvalue problems can be
applied.

We discuss two different series of Markov partitions. For the simplest
case of the first series we briefly outline of how to perform the
calculations, and we give the results for the eigenvalues and the
diffusion coefficient. For the next parameter value of this series we
give only the main results, before writing down the respective general
formulas for the whole series. For the second series we only deal with
the first two parameter values by giving the main results.

In each case we proceed by first depicting a box map of the full chain
of boxes with its Markov partition in a figure. We indicate of how the
respective Markov partition has been computed and give the exact value
of the slope by which it is defined. On this basis, we sketch the
corresponding transition matrix for the full chain of boxes and give
the main results for eigenvalues and diffusion coefficient.

\subsubsection{Series 1}
As has been pointed out in Section \ref{txt:mp}, a Markov partition is
defined via a generating orbit which obeys Eq.\ (\ref{eq:mcond}). By
using the notation of this equation, the first series of Markov
partitions discussed here is characterized by $\delta=0$ and the
number of iterations of the reduced map being $n=1$. The first case of
this series is obtained from the solution of this equation for the
slope $a$ being restricted between 2 and 4. The second case refers to
$4<a<6$, the general case is for solutions $2k<a<2(k+1)\:,\:k\in N$.\\

\noindent \underline{Case 1}:

This is the simplest case and corresponds to the smallest slope of
this series, as illustrated in Fig.\ \ref{fig:mpgen} (a). As one can
infer from the figure, the precise value of the slope can be computed
from the equation
\begin{equation}
1=2(1+\epsilon)\epsilon \quad 0\le \epsilon\le 1/2\quad ,
\end{equation}
where $a=2(1+\epsilon)$, which leads to the solution
\begin{equation}
a=(\sqrt{3}-1)/2\simeq 2.73205 \quad . 
\end{equation}
The partition of the full chain of boxes can be constructed by
continuing the box map of Fig.\ \ref{fig:mpgen} (a) periodically. The
corresponding transition matrix can then be obtained from this
partition as described in Section \ref{txt:mp} and reads
\begin{equation}
T(2.73205)=
\left( \begin{array}{ccccccccccc} 
1 & 0 & 1 & 0 & 0 & 0 & 0 & 0 & 0 & 0 & \cdots \\
1 & 0 & 1 & 0 & 0 & 0 & 0 & 0 & 0 & 0 & \cdots \\
1 & 0 & 1 & 0 & 1 & 0 & 0 & 0 & 0 & 0 & \cdots \\
0 & 1 & 0 & 1 & 0 & 1 & 0 & 0 & 0 & 0 & \cdots \\
0 & 0 & 0 & 1 & 0 & 1 & 0 & 0 & 0 & 0 & \cdots \\
0 & 0 & 0 & 1 & 0 & 1 & 0 & 1 & 0 & 0 & \cdots \\
0 & 0 & 0 & 0 & 1 & 0 & 1 & 0 & 1 & 0 & \cdots \\
\vdots & & & & \vdots & & & & \vdots \\
\end{array} \right) \quad ,
\end{equation}
where the first, second, third, $\ldots$, three rows correspond to the
first, second, third, $\ldots$, box of the chain. We now reduce this
block circulant to a simple circulant. Since the Markov partition of
each box consists of three parts, we use three different symbols
$u_m^k\:,\:v_m^k\:,\:w_m^k$ as components of the eigenvectors to write
down the eigenvalue equation, Eq.\ (\ref{eq:evpgt}), of this matrix,
\begin{eqnarray}
u_m^k+w_m^k&=&\chi_m v_m^k \nonumber \\
u_m^k+w_m^k+v_m^{k+1}&=&\chi_m w_m^k \nonumber \\
v_m^k+u_m^{k+1}+w_m^{k+1}&=&\chi_m u_m^{k+1} \quad , \quad 1<k<L \quad
,
\end{eqnarray}
supplemented by respective periodic boundary conditions for the first
and the last row of the matrix. From these equations it is immediately
obtained that
\begin{eqnarray}
w_m^k&=&\frac{v_m^{k+1}+\chi_m v_m^k}{\chi_m} \quad ,\nonumber \\
u_m^{k+1}&=&\frac{v_m^k+\chi_m v_m^{k+1}}{\chi_m}
\end{eqnarray}
which leads to
\begin{equation}
v_m^{k+1}+2v_m^k+v_m^{k-1}=\chi_m^2 v_m^k
\end{equation}
with respective periodic boundary conditions. This again defines the
eigenvalue problem of a simple circulant which we can solve by the
methods used before. For the eigenvalues we obtain
\begin{equation}
\chi_m=1\pm\sqrt{1+2\cos\theta_m}\simeq
1\pm\sqrt{3}(1-\frac{\theta_m^2}{6})\quad (L\to\infty)\quad , \quad
\theta_m=\frac{2\pi m}{L}\quad , \quad 0\le m\le L-1\quad ,
\end{equation}
where the second largest eigenvalue yields a diffusion coefficient of
\begin{equation}
D(2.73205)=\frac{\sqrt{3}}{6(1+\sqrt{3})}\simeq0.10566 \quad .
\end{equation}
As pointed out above, the largest eigenvalue is again identical to the
slope of the map. It is related to an equilibrium eigenmode which is
here a periodically continued piecewise constant function, based on
the single parts of the Markov partition.\\

\noindent \underline{Case 2}:

The second case of this series is the Markov partition defined by the
respective value of the slope between 4 and 6. Its box map partition
is illustrated in Fig.\ \ref{fig:mpgen} (b) and corresponds to
$a=2+\sqrt{8}\simeq 4.82843$. The transition matrix reads
\begin{equation}
T(4.8284)=
\left( \begin{array}{ccccccccccc} 
1 & 0 & 1 & 0 & 0 & 1 & 0 & 0 & 0 & 0 & \cdots \\
1 & 0 & 1 & 0 & 0 & 1 & 0 & 0 & 0 & 0 & \cdots \\
1 & 0 & 1 & 0 & 0 & 1 & 0 & 1 & 0 & 0 & \cdots \\
1 & 0 & 0 & 1 & 0 & 1 & 0 & 0 & 1 & 0 & \cdots \\
1 & 0 & 0 & 1 & 0 & 1 & 0 & 0 & 1 & 0 & \cdots \\
1 & 0 & 0 & 1 & 0 & 1 & 0 & 0 & 1 & 0 & \cdots \\
0 & 1 & 0 & 1 & 0 & 0 & 1 & 0 & 1 & 0 & \cdots \\
0 & 0 & 0 & 1 & 0 & 0 & 1 & 0 & 1 & 0 & \cdots \\
0 & 0 & 0 & 1 & 0 & 0 & 1 & 0 & 1 & 0 & \cdots \\
0 & 0 & 0 & 0 & 1 & 0 & 1 & 0 & 0 & 1 & \cdots \\
\vdots & & & & \vdots & & & & \vdots \\
\end{array} \right) \quad . 
\end{equation}
Again, this block circulant can be reduced to a simple circulant. By
some calculations which are quite analogous to the ones of the
previous example we obtain as eigenvalues
\begin{equation}
\chi_m=1+
\cos\theta_m\pm\sqrt{(1+\cos\theta_m)^2+2\cos(2\theta_m)+2\cos(3\theta_m)}
\simeq 2-\frac{\theta_m^2}{2}\pm\sqrt{8}(1-\frac{15}{16}\theta_m^2)\quad
(L\to\infty) \quad ,
\end{equation}
$\theta_m$ and $m$ as before, with a diffusion coefficient of
\begin{equation}
D(4.8284)=\frac{1+\frac{15}{4}\sqrt{2}}{4(1+\sqrt{2})}\simeq 0.65273
\quad . 
\end{equation}

\noindent \underline{General case of this series}:

We now give the general solution for eigenvalues and diffusion
coefficient of all cases of this series of Markov partitions. Let
\begin{equation}
a(p)=2(p+1+\epsilon)\quad \mbox{with} \quad
\epsilon=\frac{1}{2}(-p-1+\sqrt{2p+2+(p+1)^2})\quad , \quad p\in N_0
\quad ,
\end{equation}
be the slope of map $\cal L$. Then $a(0)$ is the value of the slope of
case 1 and $a(1)$ is the value of the slope of case 2. Any higher
value of $a(p)\: , \: p>1\: ,$ defines a Markov partition which is of
the same type as in the two previous examples, that is, it fulfills
the general definition of this series given at the beginning. For this
series of Markov partitions the general eigenvalue problem of the
corresponding transition matrices can be solved by generalizing the
calculations above. This yields for the eigenvalues
\begin{equation}
\chi_m=1+\sum_{k=1}^p\cos(k\theta_m)\pm\sqrt{[(1+\sum_{k=1}^p\cos(k\theta_m))^2
+ 2\sum_{k=1}^{p+1}\cos((p+k)\theta_m)} \quad ,\quad 0\le m\le L-1
\quad .
\end{equation}
In the limit of $L\to\infty$, this leads to a diffusion coefficient of
\begin{equation}
D(p)=\frac{(2p^2+p)\sqrt{(1+p)(3+p)}+2p^3+17p^2+20p+ 
6}{12(\sqrt{(1+p)(3+p)}+3+p)} \quad .
\end{equation}

\subsubsection{Series 2}
By again referring to Eq.\ (\ref{eq:mcond}), and using the notation of
Section \ref{txt:mp}, a second series of Markov partitions is defined
by $\delta=1-\epsilon$ and the number of iterations of the reduced map
being $n=1$. In the following we give the main results for only the
first two cases of this series, which are based on the solution of
this equation for the slope $a$ being $2<a<4$, and $4<a<6$,
respectively.\\

\begin{figure}
\epsfxsize=10cm
\centerline{\rotate[r]{\epsfbox{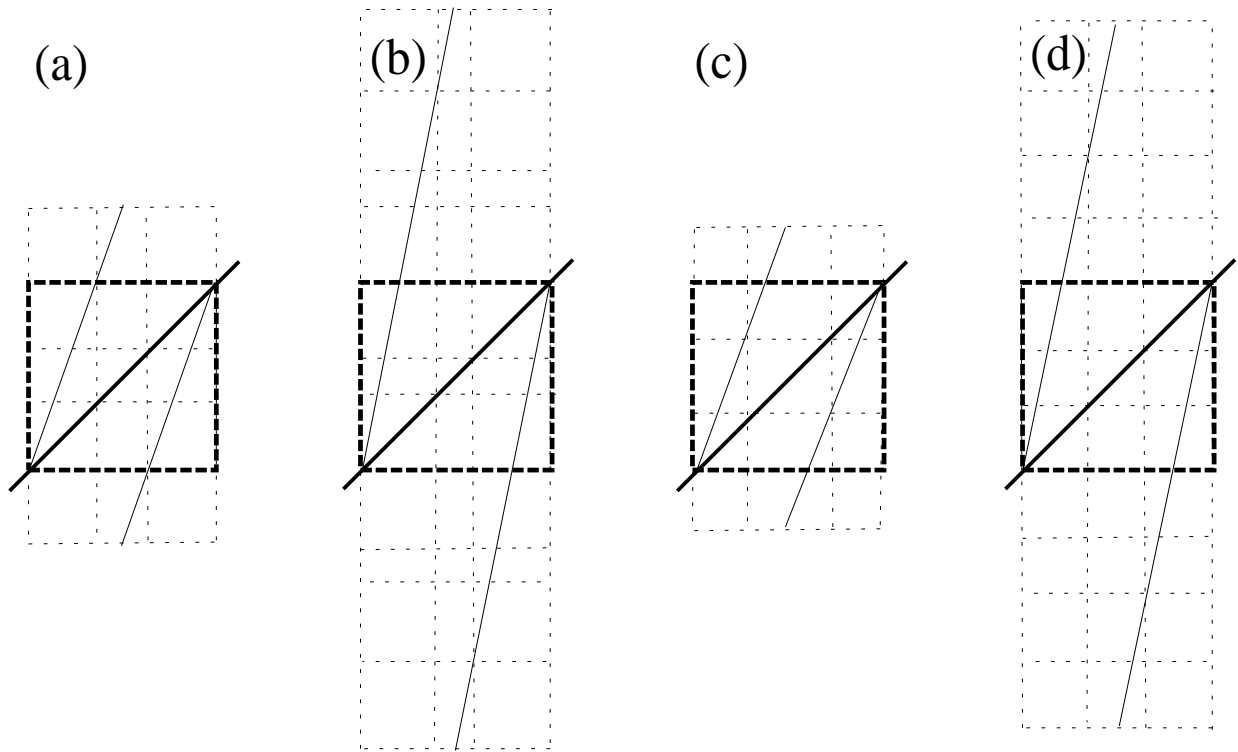}}}
 
\vspace*{0.5cm}
\caption{\label{fig:mpgen}Markov partitions for map $\cal L$ 
at values of the slope where the diffusion coefficients are computed
analytically. (a) is for slope $a\simeq 2.73205$, (b) for $a\simeq
4.82843$, (c) for $a\simeq 2.56155$ and (d) for $a\simeq 4.70156$.}
\end{figure}

\noindent \underline{Case 1}:

The partition illustrated in Fig.\ \ref{fig:mpgen} (c) corresponds to
the slope 
\begin{equation}
a=\frac{1+\sqrt{17}}{2}\simeq 2.56155
\end{equation}
and defines, periodically continued, a transition matrix of
\begin{equation}
T(2.56155)=
\left( \begin{array}{ccccccccccc} 
1 & 1 & 0 & 0 & 0 & 0 & 0 & 0 & 0 & 0 & \cdots \\
1 & 0 & 1 & 0 & 0 & 0 & 0 & 0 & 0 & 0 & \cdots \\
0 & 1 & 1 & 0 & 1 & 0 & 0 & 0 & 0 & 0 & \cdots \\
0 & 1 & 0 & 1 & 1 & 0 & 0 & 0 & 0 & 0 & \cdots \\
0 & 0 & 0 & 1 & 0 & 1 & 0 & 0 & 0 & 0 & \cdots \\
0 & 0 & 0 & 0 & 1 & 1 & 0 & 1 & 0 & 0 & \cdots \\
0 & 0 & 0 & 0 & 1 & 0 & 1 & 1 & 0 & 0 & \cdots \\
0 & 0 & 0 & 0 & 0 & 0 & 1 & 0 & 1 & 0 & \cdots \\
0 & 0 & 0 & 0 & 0 & 0 & 0 & 1 & 1 & 0 & \cdots \\
0 & 0 & 0 & 0 & 0 & 0 & 0 & 1 & 0 & 1 & \cdots \\
\vdots & & & & \vdots & & & & \vdots \\
\end{array} \right) \quad . 
\end{equation}
The eigenvalues are
\begin{equation}
\chi_m=\frac{1}{2}\pm\sqrt{\frac{9}{4}+2\cos\theta_m} \quad , \quad
\theta_m=\frac{2\pi m}{L}\quad , \quad 0\le m\le L-1\quad ,
\end{equation}
the diffusion coefficient is
\begin{equation}
D(2.56155)=\frac{2}{\sqrt{17}{1+\sqrt{17}}}\simeq 0.09468 \quad .
\end{equation}

\noindent \underline{Case 2}:

The partition illustrated in Fig.\ \ref{fig:mpgen} (d) corresponds to
the slope 
\begin{equation}
a=\frac{3+\sqrt{41}}{2}\simeq 4.70156
\end{equation}
and defines, periodically continued, a transition matrix of
\begin{equation}
T(4.70156)=
\left( \begin{array}{ccccccccccc} 
1 & 0 & 1 & 0 & 1 & 0 & 0 & 0 & 0 & 0 & \cdots \\
1 & 0 & 1 & 0 & 0 & 1 & 0 & 0 & 0 & 0 & \cdots \\
1 & 0 & 1 & 0 & 0 & 1 & 0 & 1 & 0 & 0 & \cdots \\
1 & 0 & 0 & 1 & 0 & 1 & 0 & 1 & 0 & 0 & \cdots \\
1 & 0 & 0 & 1 & 0 & 1 & 0 & 0 & 1 & 0 & \cdots \\
0 & 1 & 0 & 1 & 0 & 1 & 0 & 0 & 1 & 0 & \cdots \\
0 & 1 & 0 & 1 & 0 & 0 & 1 & 0 & 1 & 0 & \cdots \\
0 & 0 & 0 & 1 & 0 & 0 & 1 & 0 & 1 & 0 & \cdots \\
\vdots & & & & \vdots & & & & \vdots \\
\end{array} \right) \quad . 
\end{equation}
The eigenvalues are
\begin{equation}
\chi_m=\frac{1+2\cos\theta_m}{2}\pm\sqrt{\frac{1}{4}(1+2\cos\theta_m)^2+2+
2\cos\theta_m+2\cos(2\theta_m)+2\cos(3\theta_m)}\quad ,
\end{equation}
$\theta_m$ and $m$ as before, the diffusion coefficient is
\begin{equation}
D(4.70516)=\frac{1+\frac{31}{\sqrt{41}}}{3+\sqrt{41}}\simeq 0.62122
\quad . 
\end{equation}
We finally remark that analytical calculations which are similar
to the ones performed here have been carried out in Refs.\
\cite{Cvit95,Tse94,Chen95}. These calculations are
based on cycle expansions, and the diffusion coefficient has been
computed for different piecewise linear maps.
\end{appendix}

\begin{table}
\begin{tabular}{|c||c|c|c|c|c|} \hline
$i$&0&1&2&3&4\\ \hline \hline
$s_a(i)$&3.902 & 3.423 & 3.186 & 3.079 & 3.033 \\ \hline
$s_D(i)$&1.128 & 1.510 & 1.721 &
1.892 & 2.038 \\ \hline 
\end{tabular}
 
\vspace*{0.5cm}
\caption{\label{tab:scal}Scaling factors for the initial region $2\le
a\le 3$ of the parameter-dependent diffusion coefficient $D(a)$ (see
text).}
\end{table}

\begin{thebibliography}{100}
\bibitem{pili}The term {\em piecewise linear} should be understood in
the sense that the interval of the map which is periodically continued
consists of a finite number of subintervals, on the interiors of which
the map is linear.
\bibitem{rootf}Standard root-finding procedures of software packages
like NAG and IMSL require the respective functions to be
continuous. Thus, for the problem here a grid method has been
developed, where $\tilde{M}_a^n(\epsilon)-\delta$, cf.\ Eq.\
(\ref{eq:mcond}), has been discretized and where it has been checked
whether a value of this difference is close to zero by varying
$\epsilon$. Within a range of some CPU seconds of computing time on
workstations like, e.g., SUN SPARCs, a precision of up to eight digits
behind the decimal point could easily be obtained.
\bibitem{numev}Numerical evidence has been obtained from plots which
show Markov partition values of the slope with respect to their
iteration numbers $n$: It could be observed that for the three series
of Markov partition values defined above, the number of Markov
partitions in a certain range of the slope increases like a power law
with respect to the iteration number $n$ and that the Markov partition
values of the slope seem to cover the real line densely, although not
uniformly, in certain regions of the slope, with respect to increasing
the iteration number $n$.
\bibitem{negev}For example, for slope $a\simeq 6.8729834$, chainlength
$L=100$ and periodic boundary conditions, one finds numerically a
considerable number of eigenvalues less than zero, and the smallest
numerical eigenvalue is less than $-1$, whereas analytical solutions
give only a few eigenvalues less than zero, and the smallest one is
only slightly below zero and clearly greater than $-1$. For the
computations discussed here, the versions IMSL 1.1 and 2.0 and NAG
MK14B and 15A have been used.
\bibitem{tslim}The more precise values are
$a\simeq2.41421/D(a)\simeq0.10358 \: , \:
a\simeq2.41645/D(a)\simeq0.10493$; and for a maximum of second order,
e.g., $a\simeq2.20557/D(a)\simeq0.04737 \: , \:
a\simeq2.20724/D(a)\simeq0.04744$.
\end{thebibliography}
\end{document}